\documentclass[a4paper,11pt]{article}
\pdfoutput=1 

\usepackage{jheppub} 

\usepackage[T1]{fontenc} 
\usepackage{amsmath,mathtools}
\usepackage{amsthm}
\usepackage{amssymb}
\usepackage[bbgreekl]{mathbbol}
\usepackage{braket}
\usepackage{array}
\usepackage{graphicx}
\usepackage[utf8]{inputenc}
\usepackage[T1]{fontenc}
\usepackage{lmodern}
\usepackage{tensor}
\usepackage{youngtab}
\usepackage{ytableau}
\usepackage{tikz-cd} 
\usepackage{color}
\usepackage{epic}
\usepackage{pict2e}
\usepackage{bbm}

\makeatletter
\newcommand{\rref}[1]{\protected@edef\@currentlabel{#1}}
\makeatother

\DeclareMathAlphabet{\mathpzc}{OT1}{pzc}{m}{it}

\theoremstyle{definition}

 \numberwithin{equation}{section}
 \numberwithin{mdef}{section}

\let\emptyset\varnothing

\DeclareSymbolFontAlphabet{\mathbbm}{bbold}
\DeclareSymbolFontAlphabet{\mathbb}{AMSb}

\newcommand{\ZZ}{\mathbb{Z}}

\newcommand{\CC}{\mathbb{C}}
\newcommand{\lN}{\mathcal{N}}

\newcommand{\lO}{\mathcal{O}}
\newcommand{\lH}{\mathcal{H}}
\newcommand{\lS}{\mathcal{S}}
\newcommand{\lR}{\mathcal{R}}

\newcommand{\lV}{\mathcal{V}}
\newcommand{\lA}{\mathcal{A}}
\newcommand{\CL}{\mathcal{L}}

\newcommand{\lB}{\mathcal{B}}

\newcommand{\gla}{\mathfrak{gl}}

\newcommand{\GL}{\mathsf{GL}}

\newcommand{\GT}{\mathsf{GT}}
\newcommand{\GP}{\mathsf{GP}}

\newcommand{\bT}{\textbf{T}}

\newcommand{\lY}{\mathcal{Y}}
\newcommand{\gl}{\mathfrak{gl}}
\newcommand{\wT}{\mathbb{T}}

\newcommand{\bB}{\textbf{B}}
\newcommand{\bC}{\textbf{C}}
\newcommand{\T}{\mathbb{T}}

\newcommand{\Q}{\mathbb{Q}}

\newcommand{\CH}{{\mathcal{H}}}

\newcommand{\es}{{\emptyset}}
\newcommand{\svx}{{\mathsf{x}}}
\newcommand{\svX}{{\mathsf{X}}}
\newcommand{\svY}{{\mathsf{Y}}}
\newcommand{\brax}{\bra{\mathsf{x}}}

\newcommand{\bmu}{{\bar\mu}}

\newcommand{\eE}{{\rm E}}
\newcommand{\gn}{\mathsf{n}}
\newcommand{\gm}{\mathsf{m}}
\newcommand{\gln}{\gla(\mathsf{n})}

\newcommand{\YT}{{\mathcal{T}}}
\newcommand{\QEV}{{\mathbbm{\Lambda}}}
\newcommand{\gS}{{\bf S}}
\newcommand{\hq}{{\mathbbm{q}}}
\newcommand{\hhq}{{\hat{\mathbbm{q}}}}
\newcommand{\comment}[1]{}
\newcommand{\gloE}{{\mathcal{E}}}
\newcommand{\rhs}{\mbox{r.h.s.} }
\newcommand{\lhs}{\mbox{l.h.s.} }

\newcommand{\be}{\begin{eqnarray}}
\newcommand{\ee}{\end{eqnarray}}

    \usepackage{etoolbox}
    \makeatletter
    \patchcmd{\maketitle}{\@fpheader}{}{}{}
    \makeatother

\title{Separation of variables for rational $\gl(\gn)$ spin chains in any compact representation, via fusion, embedding morphism and B{\"a}cklund flow }

\author{Paul Ryan$^{a,b}$}
\author{Dmytro Volin$^{b,c}$}

\affiliation[a]{School of Mathematics \& Hamilton Mathematics Institute,\\Trinity College Dublin, College Green, Dublin 2, Ireland}
\affiliation[b]{Nordita, KTH Royal Institute of Technology and Stockholm University,\\
Roslagstullsbacken 23, SE-106 91 Stockholm, Sweden}
\affiliation[c]{Department of Physics and Astronomy,\\ Uppsala University, Box 516, SE-751 20 Uppsala, Sweden}

\emailAdd{pryan@maths.tcd.ie}
\emailAdd{dmytro.volin@physics.uu.se}

\abstract{We propose a way to separate variables in a rational integrable $\mathfrak{gl}(\gn)$ spin chain with an arbitrary finite-dimensional irreducible representation at each site and with generic twisted periodic boundary conditions. Firstly, we construct a basis that diagonalises a higher-rank version of the Sklyanin $\bB$-operator; the construction is based on recursive usage of an embedding of a $\gl(k)$ spin chain into a $\gl(k+1)$ spin chain which is induced from a Yangian homomorphism and controlled by dual diagonals of Gelfand-Tsetlin patterns. Then, we show that the same basis can be equivalently constructed by action of B{\"a}cklund-transformed fused transfer matricies, whence the Bethe wave functions factorise into a product of ascending Slater determinants in Baxter Q-functions. Finally, we construct raising and lowering operators -- the conjugate momenta -- as normal-ordered Wronskian expressions in Baxter Q-operators evaluated at zeros of $\bB$ -- the separated variables. It is an immediate consequence of the proposed construction that the Bethe algebra comprises the maximal possible number of mutually commuting charges  -- a necessary property for Bethe equations to be complete.}

\begin{document} 

\maketitle
\flushbottom
\newpage
\section{Motivation \& results}
Recently there has been substantial progress in developing the separation of variables (SoV) program for higher-rank rational $\gln$ spin chains. Initially pioneered in the works of Sklyanin \cite{10.1007/3-540-15213-X_80,Sklyanin:1991ss} for the $\gl(2)$ case and in analogy with classical integrable systems, the SoV program aims to facilitate the solution of an integrable system by reducing it to a set of decoupled "one-dimensional" problems which also implies factorisation of the wave functions. 

An important part of Sklyanin's approach relies on the so-called $\bB$-operator as we review in Section~\ref{sec:bOperator}. It is a specific polynomial in $u$ which was constructed for systems of arbitrarily high rank in \cite{2001math.ph...9013S} but its connection to the factorisation of wave functions remained unclear for quite a while until the recent findings of \cite{Gromov:2016itr,Maillet:2018bim,Ryan:2018fyo}.

One of the motivations to study higher-rank systems comes from the AdS/CFT integrable system of $\lN\!=\!4$ SYM which has a high-rank superconformal algebra $\mathfrak{psu}(2,2|4)$ as a symmetry. The quantum spectral curve \cite{Gromov:2013pga,Gromov:2014caa} which encodes the AdS/CFT spectrum is a set of equations on Baxter Q-functions which one believes to be the "one-dimensional" wave functions in a suitable SoV basis as is the case for spin chains. One hopes that their usage would substantially simplify the structure of correlation functions, as was already demonstrated in one special example \cite{Cavaglia:2018lxi}. SoV techniques of \cite{Derkachov:2001yn,Derkachov:2002tf} and their generalisations were recently employed in the computation of fishnet-type diagrams \cite{Derkachov:2018rot,Basso:2019xay,Derkachov:2019tzo}, and one expects that SoV will play an important role in further studies of the dual fishchain theories \cite{Gromov:2019aku,Gromov:2019bsj,Gromov:2019jfh}.

While unitary representations of the conformal algebra are non-compact and, moreover, the algebra is supersymmetric when applied to undeformed $\lN\!=\!4$ SYM, it was shown in \cite{Gunaydin:2017lhg,Marboe:2017dmb} that certain features of such representations can be mapped to those of compact $\gln$ representations if $\gn$ is large enough. This map requires considering spin chains in representations beyond the defining (vector) representation of $\gln$. Moreover, it was clarified in \cite{Ryan:2018fyo} how considering arbitrary $\gln$ representations facilitates understanding the regular structure of the SoV spectrum. These recent developments motivate us  to further consider spin chains in arbitrary compact representations of $\gln$, in addition to the obvious fundamental nature of the study of quantum integrability and representation theory itself.

\  \newline
In this paper we continue our analysis \cite{Ryan:2018fyo} of the interplay between the SoV $\bB$-operator \cite{Sklyanin:1992sm,2001math.ph...9013S,Gromov:2016itr} and the idea of an SoV basis construction proposed in \cite{Maillet:2018bim}. Our main result is the construction of an SoV basis for inhomogeneous $\gl(\gn)$ spin chains with any finite-dimensional irrep of $\gln$ at each local spin site and with periodic boundary conditions twisted by a matrix $G$. This basis factorises the Bethe algebra wave functions $\Psi(\svx)$ into a product of Slater determinants 
\begin{equation}\label{wavefn}
\Psi(\svx)=\braket{\svx|\Psi}=\prod_{\alpha=1}^L \prod_{k=1}^{\gn-1}\det_{1\leq i,j\leq k}\hat q_i(\svx^\alpha_{kj})\,.
\end{equation}
Here $\svx_{kj}^{\alpha}$ are  eigenvalues of the separated variables $\svX_{kj}^{\alpha}$ -- the operatorial zeros of $\bB(u)$, $\bB(\svX_{kj}^{\alpha})=0$, and $\hat q_{i}$ are eigenvalues of the Baxter operators $\hhq_i$ acting on a Bethe algebra eigenstate $\ket{\Psi}$. Analytically, $\hat q_{i}(u)=z_i^{u/\hbar}(u^{M_i}+\ldots)$ are twisted polynomials in the spectral parameter $u$ of degrees $M_i$ that depend on a chosen state $\ket{\Psi}$; and $z_1,\ldots, z_\gn$ are eigenvalues of the spin chain twist matrix $G$.

The proposed SoV basis comprises eigenvectors of the $\bB$-operator that are constructed by action of fused transfer matrices on a suitable reference state $\bra{0}$. When the $\alpha$-th spin chain site carries the highest-weight representation $\nu^\alpha=(\nu_1^\alpha,\dots,\nu^\alpha_\gn)$, we find 
\begin{equation}\label{sovbasis}
\bra{\Lambda^{\bf B}}=\bra{0}\displaystyle\prod_{\alpha=1}^L\prod_{k=1}^{\gn-1}\frac{\T_{{\bar\mu}^\alpha_k}(\theta_\alpha+\hbar\,\nu^\alpha_\gn)}{\T_{k,\nu_{k+1}^{\alpha}}(\theta_\alpha+\hbar\,\nu^\alpha_\gn)}\,,
\end{equation}
where $\bra{\Lambda^{\bf B}}$ differs from $\brax$ by a rescaling defined in \eqref{normbasis2}. In \eqref{sovbasis},  $\theta_\alpha$ are the spin chain inhomogeneities, and $\T_{{\bar\mu}^\alpha_k}$ is the transfer matrix in the representation ${\bar\mu}^\alpha_k$. ${\bar\mu}^\alpha_k=(\bar\mu_{k1}^{\alpha},\ldots,\bar\mu_{kk}^{\alpha})$ is an integer partition with $k$ components that satisfies certain constraints and relates to separated coordinates as $\svx_{kj}^{\alpha}=\theta_{\alpha}+\hbar (\mu^\alpha_{kj}+1-j)$, where $\mu^\alpha_{kj}=\bar\mu_{kj}^{\alpha}+\nu^\alpha_{k+1}$. $\T_{k,\nu_{k+1}}$ is the transfer matrix in the representation $(\nu_{k+1}^k)$, where the partition $(\nu_{k+1}^k)$ is graphically the rectangular Young diagram of size $k\times \nu_{k+1}$. For the case of rectangular representations $(S^A)$ construction \eqref{sovbasis} can be shown to be the same as in \cite{Ryan:2018fyo}. The case of the defining representation of $\lY(\gl(n))$ was also covered in \cite{Maillet:2018czd}, and of symmetric representations $(S^1)$ of $\lY(\gl(2))$ in \cite{Maillet:2019nsy}.

Unlike the SoV bases previously appearing in the literature, we construct the basis not just by action of transfer matricies but also by their inverses. While initially seeming like a complication, the action by fractions has a remarkable meaning. We find that the above ratios of transfer matrices evaluated at the inhomogeneities coincide precisely with auxiliary transfer matrices arising in the B{\"a}cklund flow procedure \cite{Krichever:1996qd,Kazakov:2007fy,Zabrodin:2007rq,Kazakov:2010iu}. Utilising this technology allows us to rewrite the SoV basis as
\begin{equation}\label{B{\"a}cklundbasis}
\bra{\Lambda^{\bf B}}=\bra{0}\prod_{\alpha=1}^L \prod_{k=1}^{\gn-1} \T^{(k)}_{\bar\mu^\alpha_k}(\theta_\alpha+\hbar \,\nu^\alpha_{k+1})\,,
\end{equation}
where $\T^{(k)}_{\mu^\alpha_k}(u)$ is a transfer matrix defined on the $\GL(k)$ strip obtained by performing a B{\"a}cklund flow $\GL(\gn)\rightarrow \GL(n-1)\rightarrow \dots \rightarrow \GL(k)$.

One can now apply the Wronskian solution 
\begin{equation}
\label{Wrointro}
\T^{(k)}_\xi(u)=\frac{\displaystyle \det_{1\leq i,j\leq k} \Q_i^{[2\hat{\xi}_j]}(u)}{\Q_{12\dots k}(u)}\,
\end{equation}
to \eqref{B{\"a}cklundbasis} to evaluate the overlap $\braket{\svx|\Psi}$ and derive \eqref{wavefn}, for appropriately normalised $\ket{\Psi}$. In \eqref{Wrointro} we used the following notations: $f^{[2n]}(u):=f(u+n\hbar)$ denotes shifts of the spectral parameter, $\hat{\xi}_j:=\xi_j-j+1$ are the shifted weights,  $\Q_{12\dots k}=\det\limits_{1\leq i,j\leq k}\Q_i^{[2(1-j)]}$, and $\Q_i$ are Baxter operators that are related to $\hhq_i$ via a gauge transformation \eqref{gaugetr}.
\newline
\newline
It is a simple consequence of the above-mentioned results that the eigenvectors $\ket{\Psi}$ with the required normalisation are built using separated variables
\be\label{Psiconstruction}
\ket{\Psi}=\prod_{\alpha=1}^L \prod_{k=1}^{\gn-1}\det_{1\leq i,j\leq k}\hat q_i(\svX^\alpha_{kj})\ket{\Omega}\,,
\ee
where $\ket\Omega$ is the unique reference state selected by the condition $\braket{\svx|\Omega}=1$.  If we choose twisted polynomials $\hat q_i$, $i=1,\ldots,\gn$ that are not eigenvalues of the operators $\hhq_i$, the construction \eqref{Psiconstruction} would be a sensible definition of off-shell Bethe states correlating with the developed SoV paradigm.
\newline
\newline
In \cite{Ryan:2018fyo} we noticed a remarkable relation between the $\bB$-operator and the so-called Gelfand-Tsetlin subalgebra of the Yangian $\lY(\gln)$ \cite{Molev1994}. Specifically, when the spin chain twist $G$ is taken to be the companion twist matrix the $\bB$-operator attains the form 
\begin{equation}\label{bisgt}
\bB(u)= \GT_1(u)\GT_2^{[2]}(u)\dots \GT_{\gn-1}^{[2(\gn-2)]}(u)+\text{nilpotent}\,.
\end{equation}
The operators $\GT_a(u)$ denote the generators of the Gelfand-Tsetlin subalgebra of the Yangian which is a maximal commutative subalgebra with several nice properties. In particular, its generators are diagonalised in the so-called Gelfand-Tsetlin basis with non-degenerate spectrum and their eigenvalues can be labelled by arrays known as Gelfand-Tsetlin patterns. On the other hand, "nilpotent" refers to a term which is strictly-upper triangular in the properly ordered Gelfand-Tsetlin basis, and hence the eigenvalues of $\bB(u)$ coincide with the eigenvalues of the above product of Gelfand-Tsetlin generators. 

In the present work we further probe this relation generalising the study from rectangular representations addressed in \cite{Ryan:2018fyo} to arbitrary finite-dimensional irreps of $\gl(\gn)$. For this generalised set up, we  prove that the $\bB$-operator is diagonalisable with $\brax$ being its eigenvectors.

There are  important technical improvements compared to \cite{Ryan:2018fyo} to cope with degeneracies in the spectrum of $\bB$. In particular, to prove that $\brax$ do indeed form a basis for generic twist eigenvalues  and inhomogeneities, we introduce auxiliary twist parameters $w_1,\ldots,w_{\gn-1}$ and show that this $(\gn-1)$-parametric deformation continuously relates the Gelfand-Tsetlin basis with the basis of $\brax$. Furthermore, we devise a sequence of embedding morphisms from lower-rank spin chains to the larger-rank spin chains pertinent to diagonalisation of $\bB$.

Finally, let us point out that we do not rely on any statements about completeness of Bethe equations. In fact, the situation is quite the opposite one --  an important ingredient of completeness theorems follows immediately from the proposed construction. Namely one shows that the Bethe algebra is a maximal commutative subalgebra of the algebra of the endomorphisms of the spin chain's Hilbert space. Indeed, the SoV basis is generated by action of transfer matrices, but it would be impossible to generate a basis if there was an extra independent operator that commutes with the transfer matrices.

Maximality of the Bethe algebra implies that the eigenstates in the Hilbert space can be unambiguously labelled by eigenvalues of Bethe algebra generators. As we can take Q-operators as generators and zeros of the Q-operators satisfy Bethe equations, we conclude that all physical states of the spin chain are labelled, and can be distinguished, by solutions of the Bethe equations.

What is not guaranteed by the above argument is that each solution of the Bethe equations labels some physical state. This question can be resolved by explicit counting but this requires certain care, especially for spin chains in arbitrary representations that we consider, as is discussed after \eqref{quantcond}. For the case of the fundamental representation the question was resolved in various ways in the literature. We mention \cite{Maillet:2019ayx} where it was discussed for the supersymmetric $\gl(2|1)$ case in the SoV framework of the same type as considered in this paper; and \cite{2013arXiv1303.1578M} where completeness is proven for $\gl(\gn)$ spin chains with and without twist, and for any value of inhomogeneities. The results of \cite{2013arXiv1303.1578M} also generalise to the supersymmetric $\gl(\mathsf{m}|\gn)$ case \cite{Chernyak:2020lgw}. 

\paragraph{Assumptions} The results of the paper are derived under the following assumptions on the values of parameters: Inhomogeneities $\theta_1,\ldots,\theta_L$ should satisfy $\theta_{\alpha}-\theta_{\beta}\neq \hbar\,k$, for any $k\in\mathbb{Z}$ and $\alpha\neq\beta$. Our SoV basis construction holds in principle for any twist eigenvalues $z_1,\dots,z_\gn$, including the degenerate case where $z_i=z_j$ for some $i\neq j$. However, we work in a special reference frame where the spin chain twist matrix is a modification of the companion matrix with eigenvalues $z_1,\dots,z_\gn$. To be able to rotate to the frame with a diagonal twist one should impose that $z_i\neq z_j$ for $i\neq j$. Aside from the mentioned restrictions, $\theta_\alpha$ and $z_j$ can be arbitrary. Modification of the companion matrix depends on the auxiliary twist parameters $w_1,\dots,w_{\gn-1}$ and these ones should be assumed to be in generic position. This generic position assumption does not affect statements that depend only on the twist eigenvalues such as the conclusion about maximality of the Bethe algebra.

\paragraph{Structure of the paper} The rest of this paper is organised as follows. In Section \ref{yangian} we review some aspects of the Yangian algebra and its representations as well as the Bethe algebra and the twists we will use. In Section \ref{gelfand} we review the Gelfand-Tsetlin algebra, introduce the embedding morphism and use it to generate the Gelfand-Tsetlin eigenvectors. In Section \ref{bsection} we discuss some properties of the $\bB$-operator and use the embedding morphism to prove that it is diagonalisable by explicitly constructing a maximal linearly independent set of its eigenvectors which deform the Gelfand-Tsetlin eigenvectors. In Section \ref{sovsection} we show that the constructed $\bB$-eigenvectors do indeed constitute a separated variable basis by demonstrating that they can be constructed by action of the Bethe algebra. We write down the Bethe wave functions in the SoV basis, and use the obtained results to construct canonically conjugate momentum operators. In the appendices we prove some technical results. 
\section{$\gln$ spin chain}\label{yangian}
\subsection{Yangian $\lY(\gln)$}
The algebraic structure underlying a rational $\gln$ spin chain is the Yangian algebra $\lY(\gln)$. $\lY(\gln)$ is the associative unital algebra with generators $T_{ij}(u),\ i,j=1,2,\dots,\gn$ subject to the RTT relation
\begin{equation}\label{rtt}
(u-v)[T_{ij}(u),T_{kl}(v)]=\hbar \left(T_{kj}(u)T_{il}(v)-T_{kj}(v)T_{il}(u) \right)
\end{equation}
for some arbitrary fixed $\hbar \in\CC^{\times}$.

The RTT relation can be conveniently by introducing two copies of $\CC^{\gn}$, referred to as auxiliary spaces, and labelled as $a$ and $b$. We then construct the triple tensor product 
\begin{equation}
{\rm End}\left(\CC^\gn\right)\otimes{\rm End}\left(\CC^\gn\right)\otimes \lY(\gl(\gn))\,.
\end{equation}
Next, we define the monodromy matrix $T(u)$ as
\begin{equation}
T(u)=\displaystyle\sum_{i,j=1}^{\gn}\eE_{ij}\otimes T_{ij}(u)
\end{equation}
where $\eE_{ij}$ are the usual basis elements of ${\rm End}\left(\CC^{\gn}\right)$ with $1$ in position $(i,j)$ and $0$ everywhere else. The RTT relation is then the statement that 
\begin{equation}
R_{ab}(u-v)T_a(u)T_b(v)=T_b(v)T_a(u)R_{ab}(u-v)
\end{equation}
where $R_{ab}(u)$ denotes the $R$-matrix 
\begin{equation}\label{YangR}
R_{ab}(u)=u\,1_{ab} - \hbar\,P _{ab}
\end{equation}
where $P_{ab}$ denotes the permutation operator on the two auxiliary spaces and $T_a(u)$ or $T_b(u)$ denotes which of the two auxiliary spaces the monodromy matrix is acting on.

Representations of $\lY(\gl(\gn))$ on some Hilbert space $\lH$ define quantum integrable models. One constructs them starting from the Lax matrix $\CL^\nu(u)$ defined by
\begin{equation}
\CL^\nu(u)=u - \hbar\,P^\nu,\quad P^\nu:=\sum_{i,j=1}^\gn \eE_{ij}\otimes \pi^\nu(\eE_{ji}),
\end{equation}
where $\nu$ is some Young diagram $\nu=(\nu_1,\dots,\nu_\gn)$ labelling a finite-dimensional irrep $\lV^\nu$ of $\gln$ and $\pi^\nu$ maps the fundamental representation generators $\eE_{ij}$ to this irrep. Then one takes 
\begin{equation}\label{spinchain}
T(u)=\CL_{L}^{\nu^L}(u-\theta_L)\dots \CL_{2}^{\nu^2}(u-\theta_2)\CL_{1}^{\nu^1}(u-\theta_1)\in {\rm End}(\CC^n\otimes \lH),
\end{equation}
with the full Hilbert space $\lH$ being a product $\lH=\bigotimes_{\alpha=1}^L \lV^{\nu^\alpha}$ of the representations $\lV^{\nu^\alpha}$ of the $\alpha$-th spin chain site.  Here $\CL_\alpha^{\nu^\alpha}$ acts non-trivially on $\CC^\gn\otimes \lV^{\nu^{\alpha}}$ and trivially on the other components of the tensor product. Note that
\be
T_{ij}(u)=\delta_{ij}u^L-u^{L-1}\left(\delta_{ij}\sum_{\alpha=1}^L\theta_L+\hbar\,\gloE_{ji}\right)+\ldots\,,
\ee
where $\gloE_{ij}=\sum_{\alpha}\pi^{\nu^{\alpha}}({\rm E}_{ij})$ are the generators of the global $\gl(\gn)$ action on the spin chain.

The parameters $\theta_\alpha\in\CC$ are known as the spin chain inhomogeneities and we impose the genericness condition
\begin{equation}\label{thetageneric}
\theta_\alpha-\theta_\beta\notin \hbar \ZZ
\end{equation}
for pairwise distinct $\alpha,\beta=1,2,\dots,L$ which is required for the spectrum of both the separated variables and the Gelfand-Tsetlin algebra to be non-degenerate. 
\newline
\newline
A useful feature of $\CL^\nu$ is its $\gln$-invariance 
\begin{equation}
[\CL^\nu(u),\eE_{ij}\otimes 1 + 1 \otimes \pi^\nu(\eE_{ij})]=0
\end{equation}
which further implies a $\GL(\gn)$ symmetry $[\CL^{\nu^{\alpha}}(u),K\otimes \Pi^{\nu^{\alpha}}(K)]=0,\ K\in \GL(\gn)$, where $\Pi^{\nu^{\alpha}}$ denotes the representation of $\GL(\gn)$ corresponding to $\pi^{\nu^\alpha}$ on $\gln$. 
This property further extends to the monodromy matrix $T(u)$:
\begin{equation}
\Pi^\nu(K)T(u)\Pi^\nu(K)^{-1}=K^{-1}T(u)K,\quad \Pi^\nu(K):=\Pi^{\nu^L}(K)\otimes \dots\otimes \Pi^{\nu^1}(K)\,.
\end{equation}
In other words, applying the same $\GL(\gn)$ transformation to each spin chain site is equivalent to performing the inverse transformation on the auxiliary space $\CC^\gn$. 
\subsection{Bethe algebra}
The Bethe algebra is the algebra of integrals of motion of the XXX chain comprising transfer matricies $\T_\xi(u)$ labelled by Young diagrams $\xi$. More precisely, the transfer matricies define a commutative family of operators
\begin{equation}\label{transfer}
[\T_\xi(u),\T_{\xi'}(v)]=0
\end{equation}
which are polynomials in the spectral parameter $u$ and the coefficients of these polynomials generate the Bethe algebra.
\newline
\newline
Not all $\T_\xi(u)$ are independent however. An independent set of generators for the Bethe algebra can be obtained from the Talalaev formula\footnote{To our knowledge, the power of this simple formula was recognised for the first time in \cite{Talalaev:2004qi}. Another related fundamental result, that Baxter $Q$-functions satisfy $\sum_{a=0}^\gn(-1)^a \T_{a,1}(u)e^{-a\hbar \partial_u}Q^{[2]}=0$, was identified earlier \cite{Krichever:1996qd}.} \cite{Talalaev:2004qi}
\begin{equation}\label{talalaev}
\det(1-T(u)e^{-\hbar \partial_u})=\displaystyle\sum_{a=0}^\gn(-1)^a \T_{a,1}(u)e^{-a\hbar \partial_u}\,,
\end{equation}
where $\T_{a,1}(u)$ denotes the transfer matrix corresponding to the Young diagram consisting of a single column with $a$ boxes.
Concretely, 
\begin{equation}
\T_{a,1}(u)=\sum_{1\leq i_1<i_2<\dots<i_a\leq \gn}T\left[^{i_1i_2\dots i_a}_{i_1 i_2\dots i_a} \right](u)\,.
\end{equation}
Here $T\left[^{i_1i_2\dots i_a}_{i_1 i_2\dots i_a} \right](u)$ are \textit{quantum minors}, defined by
\begin{equation}\label{quantumminor}
T\left[^{i_1i_2\dots i_a}_{j_1 j_2\dots j_a} \right](u)=\displaystyle\sum_{\sigma\in \gS_a}T_{i_{\sigma(1)}j_1}(u)T_{i_{\sigma(2)}j_2}^{[-2]}(u)\dots T_{i_{\sigma(a)}j_a}^{[-2(a-1)]}(u)
\end{equation}
where $\gS_a$ denotes the symmetric group on $a$ letters. 

All transfer matricies $\T_\xi(u)$ can be constructed using the fusion procedure \cite{Zabrodin:1996vm} and can be expressed in terms of $\T_{a,1}$ by means of the Cherednik-Bazhanov-Reshetikhin (CBR) formula \cite{Bazhanov:1989yk,Cherednik}
\begin{equation}\label{cbr}
\T_\xi(u)=\det_{1\leq i,j\leq \xi_1}\T_{\xi^{\rm T}_j+i-j,1}(u+\hbar(i-1))\,,
\end{equation}
where $\xi^{\rm T}$ denotes the transpose of $\xi$.
\subsection{Twist}
The spectrum of transfer matricies as constructed above is degenerate. For example for $L=1$ all transfer matricies are central elements of $U(\gln)$ and so they are scalar multiples of the identity operator acting on the spin chain. In order to remove these degeneracies it is convenient to twist by a matrix $G\in\GL(\gn)$. More precisely, one constructs the twisted monodromy matrix $\bT(u)$ defined by the replacement 
\begin{equation}
T(u)\rightarrow \bT(u):=T(u)G\,.
\end{equation}
While twisting does not define a homomorphism of the Yangian algebra since it maps the identity to $G$, it does preserve the commutation relation \eqref{rtt} due to the $\GL(\gn)$ invariance of the $R$-matrix \eqref{YangR}
\begin{equation}
[R_{ab}(u),G\otimes G]=0,\quad G\in\GL(\gn)
\end{equation}
and hence algebraic relations such as \eqref{transfer} and \eqref{cbr} are unchanged by twisting. From now on we will take all transfer matricies $\T_\xi$ to be constructed with $\bT$ instead of $T$. 
\newline
\newline 
In this paper we shall consider the case when $G$ is diagonalisable with pairwise distinct eigenvalues $z_1,z_2,\dots,z_{\gn}$. Note that by the $\GL(\gn)$ symmetry of $T(u)$ the twisted Bethe algebra is only sensitive to the twist eigenvalues -- the transfer matrix with twist $g={\rm diag}(z_1,\dots,z_{\gn})$ can be brought to the transfer matrix with any twist $G$ similar to $g$ by a simple basis change.
\newline
\newline
In our previous work \cite{Ryan:2018fyo} it proved very fruitful (from the perspective of computation simplicity when constructing an SoV basis) to consider the case where $G$ is the so-called \textit{companion twist} matrix with the eigenvalues $z_1,\dots,z_{\gn}$
\begin{equation}
G_{ij}=(-1)^{j-1}\chi_j \delta_{i1}+\delta_{i,j+1}\,,
\end{equation}
where $\chi_j$ are elementary symmetric polynomials in $z_1,\ldots,z_{\gn}$. In the present work we find it useful to introduce a generalisation of the above twist which we call the modified companion twist (MCT). It features new parameters $w_1,w_2,\dots,w_{\gn-1}$. Specifically, we have
\begin{equation}\label{modcomp}
G_{ij}=\frac{\chi_j \delta_{i1}}{w_{|j-1|}}+\delta_{i,j+1}w_j,\quad w_{|j|}:=(-1)^{j}\prod_{k=1}^j w_k \,.
\end{equation}
We stress that the $w_1,\dots,w_{\gn-1}$ do not affect the eigenvalues of the twist matrix.

For demonstration purposes, we write out the the MCT matricies explicitly for $\gn=2,3,4$:
\begin{equation}
\label{eq:growth}
\left(\begin{array}{cc}
\chi_1 & -\frac{\chi_2}{w_1} \\
w_1 & 0
\end{array} \right),\quad
\left(\begin{array}{ccc}
\chi_1 & -\frac{\chi_2}{w_1} & \frac{\chi_3}{w_1 w_2} \\
w_1 & 0 & 0 \\
0 & w_2 & 0
\end{array} \right),\quad
\left(\begin{array}{cccc}
\chi_1 & -\frac{\chi_2}{w_1} & \frac{\chi_3}{w_1 w_2} & -\frac{\chi_4}{w_1 w_2 w_3} \\
w_1 & 0 & 0 & 0\\
0 & w_2 & 0 & 0\\
0 & 0 & w_3 & 0
\end{array} \right)\,.
\end{equation}

\subsection{$\bB$-operator}
\label{sec:bOperator}
To introduce one of the key objects of this paper, the $\bB$-operator, we first discuss its classical counterpart  appearing in the study of a  spectral curve $\det\left(\lambda-L(u)\right)=0$, where $L(u)$ is a classical monodromy matrix. An eigenvector $\Psi$ of $L$ is a rational function on the curve and, provided the necessary analytic conditions are met, it is fixed by position of its poles, see {\it e.g.} \cite{babelon_bernard_talon_2003}. The poles can be described by the pairs $(\lambda,u)=(e^{p_\sigma},x_\sigma)$, where $x_\sigma$ are zeros of $B(u)$ and $e^{p_\sigma}=A(x_\sigma)$. The polynomial $B(u)$ and the rational function $A(u)$ were constructed for $\gln$ systems in  \cite{Scott:1994dz,gekhtman1995}, and the expression for $B(u)$ explicitly reads
\be
B(u)=\sum L\left[^{J_1}_{i_1}\right]L\left[^{J_2}_{J_1\ i_2}\right]\dots L\left[^{J_{\gn-1}}_{J_{\gn-2}\ i_{\gn-1}}\right] {\rm v}^{i_1}\ldots {\rm v}^{i_{\gn-1}}\,,
\ee
where  $J_k:=\{j_{k1},j_{k2},\dots,j_{kk}\}$ is a multi-index, $L\left[^{J}_{J'}\right]$ denotes the corresponding minor of the matrix $L$, and $ {\rm v}$ is a reference vector that specifies a normalisation for the eigenvector $\Psi$: ${\rm v}\cdot\Psi=1$. An alternative combination of mondoromy matrices yielding the same function $B(u)$ was proposed in \cite{Chervov:2007bb}.

Remarkably, for the Poisson bracket $\{L(u)\overset{\otimes}{,}L(v)\}=[\frac{P}{u-v},L(u)\otimes L(v)]$ which is a classical counterpart of \eqref{rtt}  one can  derive $\{x_{\sigma},p_{\sigma'}\}=\delta_{\sigma\sigma'}$. Choosing this canonical set of coordinates separates the variables in the Hamilton-Jacobi equation reducing it to a set of identical one-dimensional equations
\be
\label{eq:HJ}
\det\left(e^{\frac{\partial W}{\partial x_\sigma}}-L(x_\sigma)\right)=0\,,\quad \sigma=1,\ldots, d\,,
\ee
where $d$ is the number of degrees of freedom.

An appropriate quantisation of this formalism should yield a quantum SoV which is the key idea behind Sklyanin's SoV program.

The quantum $\bB$-operator was proposed for $\gl(3)$ in \cite{Sklyanin:1992sm} and generalised to $\gln$ in \cite{2001math.ph...9013S,Gromov:2016itr}. In terms of the twisted monodromy matrix $\bT(u)=T(u)G$ and a particular choice of the reference vector ${\rm v}$, it takes the form
\begin{equation}
\bB(u)=\sum_{J_1,\dots,J_{\gn-1}}\bT\left[^{J_1}_\gn\right]\bT^{[2]}\left[^{J_2}_{J_1\ \gn}\right]\bT^{[4]}\left[^{J_3}_{J_2\ \gn}\right]\dots \bT^{[2n-4]}\left[^{J_{\gn-1}}_{J_{\gn-2}\ \gn}\right]\,,
\end{equation}
where the entries of the multiindex $J_k=\{j_{k1},j_{k2},\dots,j_{kk}\}$ are constrained to be $1\leq j_{k1}<\dots<j_{kk}\leq \gn-1$, and $k=1,\dots,\gn-1$. The sum is then over all possible values of $j_{ki}$.

In \cite{Ryan:2018fyo} we set $G$ to be the companion twist matrix and expressed the corresponding $\bB$-operator in terms of bare (untwisted) monodromy matrix elements $T_{ij}(u)$. The same computation can be repeated when we take $G$ to be the modified companion twist (MCT) \eqref{modcomp} and we obtain
\begin{equation}\label{Bincomp}
\bB(u)=\sum_{J_1,\dots,J_{\gn-1}}T\left[^{J_1}_1\right]T^{[2]}\left[^{J_2}_{1\ J_1+1}\right]T^{[4]}\left[^{J_3}_{1\ J_2+1}\right]\dots T^{[2n-4]}\left[^{J_{\gn-1}}_{1\ J_{n-2}+1}\right]w_{J_1}w_{J_2}\dots w_{J_{\gn-1}}\,,
\end{equation}
where $w_{J_k}:=\prod\limits_{i=1}^k w_{j_{ki}}$. Notice that we can obtain $\bB$ with MCT from $\bB$ with the usual companion twist by simply replacing $T_{ij}(u)\rightarrow w_i T_{ij}(u)$ which preserves the elementary RTT relation. Hence algebraic relations involving $\bB$ with the companion twist determined in \cite{Ryan:2018fyo} can be upgraded to ones with the MCT by simply performing this transformation.

One of the main goals of this paper is to diagonalise $\bB(u)$ and to demonstrate that its eigenvectors form a separated variable basis. In order to aid with this, we recall that in \cite{Ryan:2018fyo} the explicit form \eqref{Bincomp} was shown to be closely related to another maximal commutative subalgebra (different from the previously mentioned Bethe algebra) of $\lY(\gl(\gn))$. Namely, it was noticed that the leading term in \eqref{Bincomp} where $J_k=\{1,2,\dots,k\}$ for $k=1,2,\dots,\gn-1$ belongs to the Gelfand-Tsetlin subalgebra of $\lY(\gl(\gn))$ and so the remaining terms in $\bB$ can be viewed as a deformation controlled by the parameters $w_1,\dots,w_{\gn-1}$. In the next section we will recall some details of the Gelfand-Tsetlin algebra. 

\section{Gelfand-Tsetlin algebra \& embedding morphism}\label{gelfand}
As was mentioned, the structure of the SoV basis we will construct is closely related to the Gelfand-Tsetlin basis and so knowledge of the latter is crucial for what follows. In this section we will review some aspects of the GT algebra. These tools will then be used to show that $\bB$ is diagonalisable and furthermore every eigenvector $\bra{\Lambda^\bB}$ of $\bB$ can be written as 
\begin{equation}
\bra{\Lambda^\bB}=\bra{\Lambda^{\rm GT}} + \lO\left(\dots \right)\,,
\end{equation}
where $\bra{\Lambda^{\rm GT}}$ denotes an element of the Gelfand-Tsetlin basis and $ \lO\left(\dots \right)$ denotes terms which vanish in the auxiliary singular twist limit
\be
\label{eq:ASTL}
{\rm ASTL}:\quad w_1\gg w_2\gg\dots\gg w_{\gn-1}\,.
\ee
Since the eigenvectors of $\bB$ turn out to be the eigenvectors $\brax$ of separated variables, we thus obtain that the SoV basis is a continuous deformation of the Gelfand-Tsetlin basis, with deformation parameters $w_1,\dots,w_{\gn-1}$.
\subsection{Gelfand-Tsetlin algebra}
The Gelfand-Tsetlin (GT) subalgebra of $\lY(\gln)$ can be interpreted as the Bethe algebra of the $\gln$ XXX chain with the twist matrix $G={\rm diag}(z_1,z_2,\dots,z_{\gn})$ considered in the singular twist limit\footnote{not to be confused with the ASTL defined above}
\be
\label{eq:STL}
{\rm STL:} \quad z_1\gg z_2\gg\dots\gg z_{\gn}\,.
\ee 
The Gelfand-Tsetlin generators $\GT_a(u)$, $a=1,2,\dots,\gn$ are then defined as
\begin{equation}
\GT_a(u)=\lim_{z_1\gg\dots\gg z_{\gn}} \frac{\T_{a,1}(u)}{\chi_{a}}
\end{equation}
which can easily be shown to be equal to the quantum minor $T\left[^{12\dots a}_{12\dots a} \right](u)$.

$\GT_a$ are diagonalisable and their eigenstates $\bra{\Lambda^{\rm GT}}$ are labelled as follows \cite{molev2007yangians}. Each $\Lambda$ is an $L$-tuple
\begin{equation}
\Lambda=\left( \Lambda^1,\Lambda^2,\dots,\Lambda^L\right)\,,
\end{equation}
where each $\Lambda^\alpha$ is a GT pattern. Namely, it is an array 
\begin{equation}
\begin{array}{ccccccccccc}
\nu_{1}^\alpha & \ & \nu_{2}^\alpha & \ & \dots & \ & \nu_{\gn}^\alpha \\
\ & \lambda_{\gn-1,1}^\alpha & \ & \dots & \ & \lambda_{\gn-1,\gn-1}^\alpha \\
\ & \ & \dots & \ & \dots \\
\ & \ & \lambda_{21}^\alpha & \ & \lambda_{22}^\alpha \\
\ & \ & \ & \lambda_{11}^\alpha
\end{array}
\end{equation}
in which the nodes $\lambda^\alpha_{aj}\in\ZZ$ are subject to the \textit{branching rules}
\begin{equation}
\lambda^\alpha_{a+1,j}\geq \lambda_{aj}^\alpha\geq \lambda_{a+1,j+1}^\alpha,\quad a=1,2,\dots,\gn-1,\quad j=1,2,\dots,a\,,
\end{equation}
and $\nu_{j}^{\alpha}\equiv\lambda_{\gn,j}^{\alpha}$ are fixed numbers defined by the chosen representation $\nu^{\alpha}=(\nu_{1}^{\alpha},\ldots,\nu_{\gn}^{\alpha})$ at $\alpha$-th site of the spin chain.

The eigenvalues of $\GT_a$ are
\begin{equation}\label{GTspectrum}
\bra{\Lambda^{\rm GT}}\GT_a(u)=\displaystyle\prod_{\alpha=1}^L\prod_{j=1}^a(u-\theta_\alpha-\hbar(\lambda_{aj}^\alpha+a-j))\bra{\Lambda^{\rm GT}}\,.
\end{equation}
We see that $\GT_a(u)$ measures the value of the $a$-th rows of the GT patterns which make up  $\bra{\Lambda^{\rm GT}}$. This hierarchical organisation comes from the original procedure to build up GT patterns: one considers the tautological homomorphism $\phi^{\rm GT}:T_{ij}\to T_{ij}$ which, for $i,j$ being restricted to range $1,2,\ldots, a$, can be considered as an injection  of  $\lY(\gl(a))$ into {\it e.g.} $\lY(\gl(a+1))$. One then builds the ascending chain
\be
\lY(\gl(1))\xrightarrow{\phi^{\rm GT}}\ldots \lY(\gl(a))\xrightarrow{\phi^{\rm GT}} \lY(\gl(a+1))\ldots \xrightarrow{\phi^{\rm GT}} \lY(\gl(\gn))
\ee
for which $\GT_a$ are precisely the central elements (quantum determinants) of $\lY(\gl(a))$. The center of $\lY(\gl(\gn))$ acts as
\be
\bra{\Lambda^{\rm GT}}\GT_{\gn}(u)=\displaystyle\prod_{j=1}^{\gn}\nu_j(u-\hbar(\gn-j))\bra{\Lambda^{\rm GT}}\,,\quad \nu_j(u):=\prod_{\alpha=1}^{L}(u-\theta_\alpha-\hbar\ \nu^\alpha_j)\,.
\ee

For each $\GT_a$ there is also a corresponding raising operator $\GP^+_a$ and a lowering operator $\GP^-_a$ which act on the GT basis as \cite{molev2007yangians}
\begin{equation}
\bra{\Lambda^{\rm GT}}\GP^\pm_a(\theta_\alpha+\hbar(\lambda_{aj}^\alpha+a-j))\propto \bra{\Lambda\pm\delta^\alpha_{aj}{}^{\rm GT}}\,.
\end{equation}
Here $\Lambda\pm\delta^\alpha_{aj}$ denotes a GT pattern where the node $(a,j)$ of the $\alpha$-th pattern has been changed by $\pm 1$. The coefficient of proportionality is non-zero provided that the pattern $\Lambda\pm\hbar\,\delta^\alpha_{aj}$ satisfies the branching rules, {\it i.e}. corresponds to a consistent GT pattern. 

$\GP^\pm_a(u)$ can be written explicitly in terms of quantum minors. Specifically,
\begin{equation}
\GP^+_a(u)=T\left[^{12\dots a-1\ a}_{12\dots a-1 \ a+1} \right](u),\quad \GP^-_a(u)=T\left[^{12\dots a-1\ a+1}_{12\dots a-1 \ a} \right](u)\,.
\end{equation}
\paragraph{Dual diagonals} We will find it convenient to introduce an alternative labelling of the GT pattern entries, by $\mu^\alpha_{kj}$, where $\mu^\alpha_{kj}=\lambda^\alpha_{\gn-k+j-1,j}$. For example, for $\gl(4)$ we have
\begin{equation}\label{mupattern}
\begin{array}{ccccccccccc}
\nu_{1}^\alpha & \ & \nu_{2}^\alpha & \ & \nu_{3}^\alpha & \ & \nu_{4}^\alpha \\
\ & \mu_{11}^\alpha & \ & \mu_{22}^\alpha & \ & \mu_{33}^\alpha \\
\ & \ & \mu_{21}^\alpha & \ & \mu_{32}^\alpha \\
\ & \ & \ & \mu_{31}^\alpha
\end{array}\,.
\end{equation}
This new labelling naturally suggests to parameterise GT patterns by what we refer to as \textit{dual diagonals} $\mu^\alpha_k$ where we define
\begin{equation}
\mu^\alpha_k=(\mu^\alpha_{k1},\mu^\alpha_{k2},\dots,\mu^\alpha_{kk}),\quad k=1,\dots,\gn-1\,.
\end{equation}
Since the minimum value of each $\mu^\alpha_{kj}$ allowed by the branching rules is $\mu^\alpha_{kj}=\nu^\alpha_{k+1}$, it is also convenient to introduce the parameters 
\begin{equation}
\label{eq:defmbar}
\bar\mu^\alpha_{kj}=\mu^\alpha_{kj}-\nu^\alpha_{k+1}
\end{equation}
which measure how much a given dual diagonal has been excited above its minimum value. Clearly, $\bar\mu^\alpha_k$ corresponds to a $\gl(k)$ Young diagram. As we will see, dual diagonals turn out to be a natural labelling of GT patterns in the context of separation of variables.
\subsection{Embedding morphism}
\label{sec:embeddingmorphism}
As was described above, the Gelfand-Tsetlin algebra is constructed by considering the tautological injection $T_{ij}\mapsto T_{ij}$ of $\lY(\gl(k))$ into $\lY(\gl(k+1))$. Now consider a different (nearly) tautological injection of $\lY(\gl(k))$ into $\lY(\gl(k+1))$ defined by
\be
\label{embtwo}
\phi:T_{ij}(u)\mapsto T_{1+i,1+j}(u)\,.
\ee 
We use it for a different purpose: to construct a special embedding of a $\gl(k)$ spin chain into a $\gl(k+1)$ chain that shall be called embedding morphism. Formally the embedding morphism is an induced map $\phi:\CH_{k}\to \CH_{k+1}$, where $\CH_{k}$ is the Hilbert space of the $\gl(k)$ spin chain of length $L$ with spin chain sites in irreps $(\nu_1^{\alpha},\ldots,\nu_k^{\alpha})$, fully defined by the following property
\be\label{eq:embphi}
 \phi:\bra{0_k}\mathcal{J}\mapsto \bra{0_{k+1}}\phi(\mathcal{J})\,,
\ee
where $\mathcal{J}$ is any element of $\lY(\gl(k))$, and $\bra{0_k}$ is the lowest-weight vector of the $\gl(k)$ chain -- the state whose GT pattern has the lowest possible entries $\mu_{ij}^{\alpha}=\nu_{i+1}^{\alpha}$ for $i=1,2,\ldots,k-1$, $j=1,2,\ldots, i$.

Define $\lV_{(k)}:=\phi(\CH_{k})$.  By abuse of notation we may also use $\lV_{(k)}=\phi^m(\CH_{k})$, for $m=2,3,\ldots,\gn-k$ and so in particular we think about $\lV_{(k)}$ as a subspace in the full $\gln$ spin chain which represents a smaller $\gl(k)$ chain.
\newline
\newline
Remarkably, the embedding morphism has a simple coordinatisation using GT patterns:
\be
\label{eq:im1}
\phi\left(
\mbox{
\begin{picture}(72,26)(8,23)
\put(4,40){
$
\begin{array}{ccccccccccc}
\nu_{1}^\alpha & \ &  \ &  \ & \ldots & \ &  \ & \ & \nu_{k}^\alpha  
\end{array}
$
}
\put(33,20){
$\mu_{ij}^{\alpha}$
}
\drawline(16,32)(70,32)(43,5)(16,32)
\end{picture}
}
\right)
\propto
\mbox{
\begin{picture}(100,30)(8,16)
\put(5,40){
$
\begin{array}{cccccccccccc}
\nu_{1}^\alpha & \ &  \ &  \ & \ldots & \ & \ & \ & \nu_{k}^\alpha   & \ & \  & \nu_{k+1}^{\alpha}
\end{array}
$
}
\put(33,20){
$\mu_{ij}^{\alpha}$
}
\put(53,-2){$\nu_{k+1}^{\alpha}$}
\put(78,23){$\nu_{k+1}^{\alpha}$}
\put(68,9){\rotatebox{45}{$\ldots$}}
\drawline(16,32)(70,32)(43,5)(16,32)
\end{picture}
}
\,,
\ee
{\it i.e.} the image of a state with the GT pattern $\Lambda'$ for the $\gl(k)$ spin chain is the state for the $\gl(k+1)$ chain with the GT pattern which has the right-most dual diagonal at the lowest possible value and the remaining triangular block coinciding with $\Lambda'$. 

The above implies the following property of $\lH_{k+1}$ which we will frequently use. If $\bra{\Lambda}\in\lH_{k+1}$ is obtained from a vector in $\lH_{k}$ by action of $\phi$ then $T_{11}(u)=\GT_1(u)\in\lY(\gl(k+1))$ with the eigenvalue $\nu_{k+1}(u)$. Since the eigenvalue of $T_{11}$, and hence of the global Cartan generator $\gloE_{11}$, is at its lowest possible value and the eigenvalue of $\gloE_{11}$ is lowered by $T_{j1},\ j>1$ it follows that
\begin{equation}\label{lowering}
\bra{\Lambda}T_{j1}(u)=\delta_{j1}\nu_{k+1}(u)\bra{\Lambda},\quad j=1,\dots,k+1.
\end{equation}
To see why the property \eqref{eq:im1} indeed holds it is enough to check that the raising operators $\GP_a^{+}$ act accordingly because their action generates the whole Hilbert space starting from the lowest-weight state. To this end consider yet another family of homomorphisms \cite{molev2007yangians} $\psi_m:\lY(\gl(k))\longrightarrow \lY(\gl(k+m))$ for $m=1,2,\ldots$ defined by
\begin{equation}\label{psidefn}
\psi_m:T_{ij}(u)\mapsto \left(\GT_m(u+m\hbar) \right)^{-1}T\left[^{1\ldots m\ m+i}_{1\ldots m\ m+j} \right](u+m\hbar)\,.
\end{equation}
One can show that, for any quantum minor $T\left[^\lA_\lB\right](u)$,
\begin{equation}
\psi_m:T\left[^\lA_\lB\right](u)\mapsto \left(\GT_m(u+m\hbar) \right)^{-1}T\left[^{1\ldots m\ \lA+m}_{1\ldots m\ \lB+m} \right](u+m\hbar)\,,
\end{equation}
and that $\psi_m=(\psi_1)^m$. Then
\be
\label{eq:im2}
\psi_1(\GP_a^{\pm}(u))=\left(\GT_1(u+\hbar) \right)^{-1}\GP_{a+1}^{\pm}(u+\hbar)\,.
\ee
Define an embedding morphism of spin chains $\psi_1:\CH_{k}\to\CH_{k+1}$ by \eqref{eq:embphi} with $\phi$ replaced by $\psi_1$. Given \eqref{eq:im2}, relation \eqref{eq:im1} with $\phi$ replaced by $\psi_1$ is obvious: on one hand, \eqref{eq:im2} states that action of raising and lowering operators commutes, up to normalisation, with $\psi_1$. On the other hand, one gets in the image of $\psi_1$ precisely the states of $\CH_{k+1}$ that are generated by $\GP_2^+,\GP_3^+,\ldots,\GP_k^+$ acting on $\bra{0_{k+1}}$. Finally, one notes that the last dual diagonal cannot be excited by these operators if the node $\mu_{k1}^{\alpha}$ attains its lowest value $\mu_{k1}^{\alpha}=\nu_{k+1}^{\alpha}$. But $\mu_{k1}^{\alpha}$ can only change by action of $\GP_1^+$ which cannot be represented as $\psi_1(\GP_a^+)$.

Now we remark that the embeddings $\psi_1$ and $\phi$ coincide. Indeed, for any $\bra{\Lambda}$ of the $\gl(k+1)$ chain with $\mu_{k1}^{\alpha}=\nu_{k+1}^{\alpha}$ one has $\bra{\Lambda}T_{j1}(u)=\delta_{j1}\nu_{k+1}(u)\bra{\Lambda}$ as was established above, and so one computes
\be
\bra{\Lambda}\psi_1(T_{ij}(u))&=&(\nu_{k+1}(u+\hbar))^{-1}\bra{\Lambda}T\left[^{1\ 1+i}_{1\ 1+j}\right](u+\hbar)
\nonumber\\
&=&\bra{\Lambda}\phi(T_{ij}(u))\,.
\ee
Hence $\psi_1(T_{ij}(u))=\phi(T_{ij}(u))$ when restricted to $\lV_{(k)}$, and so \eqref{eq:im1} holds.

\subsection{A roadmap to the GT basis}
Finally, we present a special generation of states in the GT basis based on the embedding morphism. The idea is to consider a recursive procedure
\be
\label{eq:CHLV}
 \cdots \to\CH_{k}\lhook\joinrel\xrightarrow{\ \phi\ } \lV_{(k)} \xrightarrow{\lS} \CH_{k+1} \lhook\joinrel\xrightarrow{\ \phi\ }\cdots\,, 
 \ee
where $\lS$ is the introduced-below composite raising operator that excites the largest dual diagonal from its lowest to the desired value. The recursion starts from the lowest weight state of the $\gl(2)$ spin chain which spans $\lV_{(1)}$ and terminates with the full Hilbert space $\CH_{\gn}$.
\newline
\newline
We start by considering a state $\bra{\Lambda}\in\lH_{k+1}$ obtained from a state in $\lH_k$ by action of the embedding morphism. By definition, $\Lambda$ is an $L$-tuple of patterns $\Lambda=(\Lambda^1,\dots,\Lambda^L)$ and each $\Lambda^\alpha$ has $\mu^\alpha_{kj}=\nu^\alpha_{k+1}$, $j=1,\dots,k$. From here we will construct a state where $\mu_{kj}^\alpha=\nu^\alpha_{k+1}+1$, $j=1,\dots,a$, $\mu^\alpha_{kj}=\nu^\alpha_{k+1}$ for $j>a$, for some $1\leq a\leq k$. By the properties of the GT raising operators we know that we can obtain such a state by acting on $\bra{\Lambda}$ with the operators which raise those particular nodes, obtaining 
\begin{equation}\label{creating}
\bra{\Lambda}\GP^+_1\GP^+_2\dots\GP^+_a\,,
\end{equation} 
where each $\GP^+$ is evaluated at $\theta_\alpha+\hbar\,\nu^\alpha_{k+1}$. This can be written explicitly in terms of minors as
\begin{equation}\label{dualraising1}
\bra{\Lambda}T\left[^{1}_{2}\right]T\left[^{12}_{13}\right]\dots T\left[^{12\dots a-1\ a}_{12\dots a-1\ a+1}\right]\,.
\end{equation}
By straightforward application of the quantum column expansion of minors \cite{molev2007yangians} one can show that \eqref{dualraising1} coincides, up to a non-zero coefficient, with
\begin{equation}\label{contraction}
\bra{\Lambda}T\left[^{12\dots a}_{23\dots a+1}\right](\theta_\alpha+\hbar\,\nu^\alpha_{k+1})\,.
\end{equation}
From here, one can further excite the excited nodes, filling up a certain number of nodes successively by $1$ until the full dual diagonal has reached the desired value. In summary, we have the following.
For a Young diagram $\bmu_k$ of height $h_{\bmu_k}\leq k$, let us define a composite operator $\lS_{\bmu_k}(u)$ by
\begin{equation}\label{compositeraising}
\lS_{\bmu_k}(u)=\prod_{j\in {\rm col}(\bmu_k)}^\rightarrow \lS_{\bmu_k,j}(u+\hbar(j-1))\,,
\end{equation}
where the product is over the number of columns ${\rm col}(\bmu_k)$ of $\bmu_k$; and $\lS_{\bmu_{k,j}}$ is the raising operator associated to the $j$-th column of $\bmu_k$. Specifically, if we let $h_{\bmu_k}^j$ denote the number of boxes in the $j$-th column of $\bmu_k$ then 
\begin{equation}
S_{\bmu_k,j}(u)=T\left[^{1 \ 2\ \dots\ h^j_{\bmu_k}}_{2\ 3\ \dots\ h^j_{\bmu_k}+1}\right](u)\,.
\end{equation}
Then $\bra{\Lambda}\prod\limits_{\alpha=1}^L S_{\bmu_k^{\alpha}}(\theta_{\alpha}+\hbar\nu_{k+1}^{\alpha})$ is a state in $\CH_{k+1}$ whose $k$-th dual diagonals are excited to values $\mu_k^1,\mu_k^2,\ldots,\mu_k^L\,.$

Finally, by running the recursion \eqref{eq:CHLV}, we can write any element of the GT basis as 
\begin{equation}
\bra{\Lambda^{\rm GT}}=\bra{0}\displaystyle\prod_{k}^{\leftarrow}\prod_{\alpha=1}^L \phi^{\gn-k-1}\left(\lS_{\bmu^\alpha_{k}}(\theta_\alpha+\hbar\,\nu^\alpha_{k+1})\right)\,,
\end{equation}
where the first product ranges over $k=1,\dots,\gn-1$.
\section{Diagonalising the $\bB$-operator}\label{bsection}
As was reviewed in the introduction, the eigenvectors of separated variables have been conjectured, and proven in certain cases, to be eigenvectors of the $\bB$-operator. The most general result achieved so far was to construct \cite{Ryan:2018fyo} a family of $\bB$ eigenvectors for $\gln$ spin chains in $(S^A)$ representations. Unfortunately, for certain classes of representations the spectrum of $\bB$ is degenerate\footnote{It is non-degnerate for symmetric and antisymmetric powers of fundamental representations, their conjugates, and some other special cases.} and so linear independence of the eigenvectors constructed in \cite{Ryan:2018fyo} cannot be inferred from the eigenvalues of $\bB$ alone. Furthermore, it is not even granted that $\bB$ is diagonalisable.
\newline
\newline
In this section we present a procedure that resolves both of these issues and furthermore generalises the results of \cite{Ryan:2018fyo} to arbitrary compact representations. The idea is to construct the eigenvectors of $\bB$ by ascending through the spin chains of increasing rank
\be
\label{eq:CHLV2}
 \cdots \to\CH_{k}\lhook\joinrel\xrightarrow{\ \phi\ } \lV_{(k)} \xrightarrow{\T_{\bar\mu_k}} \CH_{k+1} \lhook\joinrel\xrightarrow{\ \phi\ }\cdots\,. 
\ee
The procedure is rooted in the following two observations. Firstly,
\be
\label{eq:bBk}
\bB^{(k+1)}{|}_{\lV_{(k)}}\sim \phi\left(\bB^{(k)}\right){|}_{\lV_{(k)}}\,,
\ee
where $\bB^{(k)}$ denotes the $\bB$-operator for the $\gl(k)$ spin chain, and $\sim$ means equality up to multiplication by an operator which is proportional to the identity when restricted to $\lV_{(k)}$. This property allows one to build all eigenstates of $\bB^{(k+1)}$ for which the last dual diagonal is not excited, simply by applying the embedding morphism to smaller-rank chains.

Secondly, we excite the last dual diagonal of $\gl(k+1)$ patterns by action of transfer matrices  $\T_{\bar\mu_k}$, where the choice of representation $\bar\mu_k$ dictates how the diagonal should be excited. This step closely follows the results of \cite{Ryan:2018fyo}.

To check that the outlined procedure does indeed produce a basis of $\CH_{\gn}$, we analyse it in the ASTL \eqref{eq:ASTL} where it degenerates to the construction \eqref{eq:CHLV} of GT eigenvectors which are known to form a basis.

\subsection{Properties of $\bB$}

From \eqref{Bincomp}, it is straightforward to deduce the decomposition \eqref{bisgt} of $\bB$ into diagonal and nilpotent upper-triangular components which we abbreviate as $\bB=\bB^{\rm GT}+{\rm Nil}$. The relative magnitude of the ${\rm Nil}$ term is controlled by auxiliary twist parameters, and we can fully suppress it by taking the ASTL \eqref{eq:ASTL}. Hence, we can perceive eigenvectors of $\bB$ as a continuous deformation of the GT eigenvectors for finite values of $w_{1},\ldots,w_{\gn-1}$ and therefore label them by the GT patterns: An eigenvector of $\bB$  is denoted by $\bra{\Lambda^{\bB}}$ if it becomes $\bra{\Lambda^{\rm GT}}$ in the ASTL. Due to degeneracy of the spectrum of $\bB^{\rm GT}$, there are legitimate questions about existence and unicity of such vectors, but we overcome these issues by explicitly building them in the next subsection. Meanwhile, the eigenvalue of  $\bB$ on $\bra{\Lambda^{\bB}}$ is guaranteed to be equal to that of $\bB^{\rm GT}$ on $\bra{\Lambda^{\rm GT}}$:
\be
\label{eq:variablesx}
\bra{\Lambda^{\bB}}\bB(u)=\prod_{\alpha=1}^{L}\prod_{k=1}^{\gn-1}\prod_{j=1}^k(u-\svx_{kj}^{\alpha})\bra{\Lambda^{\bB}}\,,\quad \svx_{kj}^{\alpha}=\theta_{\alpha}+\hbar\,(\mu_{kj}^{\alpha}-j+1)\,,
\ee
where $\mu^\alpha_{kj}$ are the entries of the GT patterns $\Lambda$ as explained in \eqref{mupattern}. $\bB(u)$ is a polynomial in $u$ of degree $L\frac{\gn(\gn-1)}{2}$ so we can write it as
\be
\bB(u)=\prod_{\alpha=1}^{L}\prod_{k=1}^{\gn-1}\prod_{j=1}^k(u-\svX_{kj}^{\alpha})\,,
\ee
and $\svX_{kj}^{\alpha}$ are defined unambiguously as the operators with eigenvalues $\svx_{kj}^{\alpha}$. They form a maximal commutative subalgebra of ${\rm End}(\CH_\gn)$ provided $\bB(u)$ is diagonalisable and its diagonalisation is performed in a $u$-independent way. This becomes clear when we construct $\bra{\Lambda^{\bB}}$ explicitly in the next section.
\newline
\newline
Let us now understand how the crucial property \eqref{eq:bBk} comes about. The \rhs of \eqref{eq:bBk} is the image of $\bB^{(k)}$, and $\bB^{(k)}$ is defined by  \eqref{Bincomp} with $\gn$ being replaced with $k$. It is an operator acting on $\CH_{k}$. The \lhs of \eqref{eq:bBk} contains the operator $\bB^{(k+1)}$ acting on $\CH_{k+1}$. We illustrate its restriction to the subspace $\lV_{(k)}$ for the case $k+1=\gn$. From  \eqref{lowering} and the definition of minors \eqref{quantumminor} it follows that $T^{[2r]}\left[^{J_{r+1}}_{1\ J_r+1}\right]$ is only non-zero if $J_{r+1}$ contains $1$. Denote then $J_{r+1}=(1\ J'_{r+1}+1)$ and then simplify, using \eqref{lowering}, $T^{[2r]}\left[^{1\ J'_{r+1}+1}_{1\ J_r+1}\right]=\nu_{\gn}(u+\hbar r)\phi\left(T^{[2(r-1)]}\left[^{J'_{r+1}}_{J_r}\right]\right)$.  Overall, one gets
\be
\label{eq:bBk2}
\bB^{(\gn)}{|}_{\lV_{(\gn-1)}}=\prod_{r=0}^{\gn-2}\nu_{\gn}(u+\hbar\,r)\, \phi\left(\bB^{(\gn-1)}\right){|}_{\lV_{(\gn-1)}}\,.
\ee
Obviously, the above conclusion holds when we replace $\gn$ with $k+1$ which confirms \eqref{eq:bBk}.
\newline
\newline
As already outlined, \eqref{eq:bBk} ensures that eigenvectors of $\bB^{(k)}$ become eigenvectors of $\bB^{(k+1)}$ upon using the embedding morphism. Moreover, one guarantees that $\bra{\Lambda^{\bB}}\in\lV_{(k)}\subset\CH_{\gn}$ if and only if at most the first $k-1$ dual diagonals  are excited above their minimal values (for each $\Lambda^\alpha$ of the pattern $\Lambda=(\Lambda^1,\ldots,\Lambda^L)$). This is not a trivial conclusion as $\bra{\Lambda^{\bB}}$ deforms $\bra{\Lambda^{\rm GT}}$ and so its relation to the subspaces $\lV_{(k)}$ could become obscured.  It allows us to consider $\svX_{k'j}^{\alpha}$ as operators defined for any $\gl(k)$ chain with $\svX_{k'j}^{\alpha}=\phi^*(\svX_{k'j}^{\alpha})$, where $\phi^*$ is a pullback of the embedding morphism. For $k>k'$, these operators, for generic representations, are dynamical having all possible eigenvalues permitted by branching rules. For $k\leq k'$,  $X_{k'j}^{\alpha}$ are non-dynamical and they attain only their lowest values.
\subsection{Building up $\bB$ eigenvectors}
In the previous subsection we clarified how the embedding $\CH_{k}\lhook\joinrel\xrightarrow{\ \phi\ } \lV_{(k)}\subset\CH_{k+1}$ works. This subsection focuses mostly on the excitation step $ \lV_{(k)} \xrightarrow{\T_{\bar\mu_k}} \CH_{k+1}$. We understand by now that one should focus on exciting the longest dual diagonal as all the other diagonals should have been excited to the desired values at lower-rank stages of the recursion.
\newline
\newline
The $\bB$-operator is independent of the twist matrix eigenvalues $z_1,\dots,z_{\gn}$ and hence so are its eigenvectors. Since we expect to construct eigenvectors of $\bB$ with transfer matricies $\T_{\xi}$, it is natural then to check the case of the null twist first, where the null twist is defined as the MCT with  $z_j=0$. In \cite{Ryan:2018fyo} we derived the following commutation relation between $\bB$ and transfer matricies $\T_{\xi}^\lN$ computed in the null twist frame:
\begin{equation}\label{BTcomm}
\T_{\xi}^\lN(v)\bB(u)=f_{\xi}(u,v)\bB(u)\T_{\xi}^\lN(v)+\lR(u,v)\,,
\end{equation}
where $f_{\xi}(u,v)$ is a function given explicitly by 
\begin{equation}
\label{fvalue}
f_{\xi}(u,v)=\displaystyle\prod_{a=1}^{h_\xi}\frac{u-v+\hbar(a-1-\xi_a)}{u-v+\hbar(a-1)}\,,
\end{equation}
and $\lR(u,v)=\sum_{j=1}^nT_{j1}(v)\times \dots$. This relation also holds when the auxiliary parameters $w_i$ are introduced, the only difference is in the rescaling by positive powers of $w_i$ of terms of $\lR$. 
\newline
\newline
Our goal is to engineer a situation when the remainder $\lR(u,v)$ vanishes. Then we can use \eqref{BTcomm} to intertwine between eigenstates of $\bB$.

 We say that $\bra{\Lambda}$ is an {\label{pos:admissible}admissible} vector at point $v$ if it is an eigenstate of $\bB$ and it satisfies $\bra{\Lambda}T_{j1}(v)=0$ for all $j$ and the given value of $v$. 

From \eqref{BTcomm}, it is clear that if $\bra{\Lambda}$ is admissible at point $v$ then $\bra{\Lambda}\T_{\xi}^\lN(v)$ is  an eigenstate of $\bB$ provided that the action of $\T^\lN_{\xi}(v)$ on $\bra{\Lambda}$ is non-zero. However, recall that we are eventually interested in action of transfer-matrices $\T_{\xi}$ with non-null twist,  and it is not obvious that $\bra{\Lambda}\T_{\xi}^\lN(v)$ coincides with $\bra{\Lambda}\T_{\xi}(v)$ under the above assumptions. To cover this point, we briefly discuss the relevant properties of transfer matricies $\T_{\xi}$, more details can be found in \cite{Ryan:2018fyo}. 

Transfer matricies $\T_{\xi}(u)$ can be obtained as the trace of the fused monodromy matrix $\bT_{\xi}$. The elements of $\bT_{\xi}(u)$ are what we refer to as $\xi$-minors $\bT_{\xi}\left[^\lA_\lB\right](u)$. For a $\gl(k+1)$ spin chain, $\lA$ and $\lB$ are sets of indices taking values $1,2,\dots,k+1$ that are in correspondence with semi-standard Young tableaux of shape $\xi$
\begin{equation}
\lA=\raisebox{-0.4\height}{
\begin{picture}(90,70)(-10,0)
\put(-8,68){\vector(0,-1){30}}
\put(-8,68){\vector(1,0){40}}
\put(-16,42){$a$}
\put(25,72){$s$}
\drawline(0,0)(0,60)
\drawline(20,0)(20,60)
\drawline(40,20)(40,60)
\drawline(60,40)(60,60)
\drawline(80,40)(80,60)
\drawline(0,0)(20,0)
\drawline(0,20)(40,20)
\drawline(0,40)(80,40)
\drawline(0,60)(80,60)
\put(2,48){$\scriptstyle {\mathcal{A}}_{1,1}$}
\put(22,48){$\scriptstyle {\mathcal{A}}_{1,2}$}
\put(44,48){$\ldots$}
\put(60.5,48){$\scriptstyle {\mathcal{A}}_{1\!,\xi_1}$}
\put(2,28){$\scriptstyle {\mathcal{A}}_{2,1}$}
\put(24,28){$\ldots$}
\put(4,8){$\ldots$}
\end{picture}
}\,,\quad
\lB=\raisebox{-0.4\height}{
\begin{picture}(90,70)(-10,0)
\drawline(0,0)(0,60)
\drawline(20,0)(20,60)
\drawline(40,20)(40,60)
\drawline(60,40)(60,60)
\drawline(80,40)(80,60)
\drawline(0,0)(20,0)
\drawline(0,20)(40,20)
\drawline(0,40)(80,40)
\drawline(0,60)(80,60)
\put(2,48){$\scriptstyle {\mathcal{B}}_{1,1}$}
\put(22,48){$\scriptstyle {\mathcal{B}}_{1,2}$}
\put(44,48){$\ldots$}
\put(60.5,48){$\scriptstyle {\mathcal{B}}_{1\!,\xi_1}$}
\put(2,28){$\scriptstyle {\mathcal{B}}_{2,1}$}
\put(24,28){$\ldots$}
\put(4,8){$\ldots$}
\end{picture}
}\,.
\end{equation}
$\bT_{\xi}\left[^\lA_\lB\right](u)$ are constructed by applying appropriate symmetrization of the indices in the ordered product $\overrightarrow{\prod\limits_{a=1}^{h_{\xi}}\prod\limits_{s=1}^{\xi_a}}\bT\left[^{\lA_{a,s}}_{\lB_{a,s}}\right](u+\hbar(s-a))$, of which \eqref{quantumminor} is an example for $\xi=(1^a)$. The transfer matrix $\T_\xi$ is then defined as $\T_\xi(u)=\sum_{\lA}\bT_{\xi}\left[^\lA_\lA\right](u)\,,$ where the sum is over all admissible tableaux $\lA$. It is then a straightforward computation to demonstrate
\begin{equation}
\label{eq:Txi}
\T_{\xi}(v)=\sum_{\lA}w_\lA T_{\xi}\left[^\lA_{\lA+1}\right](v)+\sum_{j}T_{j1}(v)\times \lO(z_1,\dots,z_{k+1})\,,
\end{equation}
where  $w_\lA:=\prod_{a\in \lA}w_a$.
\newline
\newline
The first term in \eqref{eq:Txi} coincides with $\T_{\xi}^\lN$ and we clearly see that the second term vanishes when acting on an admissible vector at point $v$ and thus indeed $\bra{\Lambda}\T_{\xi}^\lN(v)=\bra{\Lambda}\T_{\xi}(v)$. One may ask how $ z_1,\ldots  z_{k+1}$ -- the eigenvalues of the MCT of the $\gl(k+1)$ spin chain are related to $z_1,\ldots z_{\gn}$ -- the original MCT eigenvalues. The point here is that none of the constructed states depend on $z_{i}$ and so this relation is immaterial. The auxiliary parameters $w_i$ should however be compatible with the injection \eqref{embtwo} used in the embedding procedure: If $w_i^{(k)}$ denote the auxiliary parameters used for transfer matrices of $\lY(\gl(k))$ then $w_{i+1}^{(k+1)}=w^{(k)}_{i}$, $i=1,\dots,k$. 
\newline
\newline
Let $\bra{\Lambda'}$ be an eigenvector of $\bB^{(k)}$. Then we use \eqref{lowering} to readily see that $\bra{\Lambda}=\phi(\bra{\Lambda'})$ is an admissible vector at points $\theta_{\alpha}+\hbar\,\nu^{\alpha}_{k+1}$. Hence, to excite the $k$-th dual diagonals $\mu^\alpha_{kj}$ of  patterns $\Lambda^{\alpha}$, $\alpha=1,\ldots,L$ we should consider the following product 
\be
\label{actiononadmissible}
\bra{\Lambda}\prod_{\alpha=1}^L \T_{\bar\mu_{k}^{\alpha}}(\theta_{\alpha}+\hbar\,\nu^{\alpha}_{k+1})\,
\ee
as one can confirm from the explicit vale of $f_\xi(u,v)$ \eqref{fvalue} for $\xi=\bar\mu_k^{\alpha}$. The only thing to check is that the action of $\T_{\bar\mu_{k}^{\alpha}}$ at the point $(\theta_{\alpha}+\hbar\,\nu^{\alpha}_{k+1})$ on $\bra{\Lambda}$ results in a vector which is still admissible at points $(\theta_{\beta}+\hbar\,\nu^{\beta}_{k+1})$ for $\beta\neq\alpha$. This is verified by considering the following fused RTT relation \cite{Ryan:2018fyo}:
\begin{equation}\label{fusedRTT}
(v-v')[T_{j1}(v),T_{\bar\mu_k}\left[^\lA_\lB\right](v')]=\sum_{a\in \lA}T_{a1}(v)\times\dots - \sum_{a\in \lA}T_{a1}(v')\times\dots\,.
\end{equation}
Taking $v=(\theta_{\beta}+\hbar\,\nu^{\beta}_{k+1})$, $v'=(\theta_{\alpha}+\hbar\,\nu^{\alpha}_{k+1})$ and using \eqref{lowering} and \eqref{fusedRTT} we conclude that if $\bra{\Lambda}$ is admissible at points $v,v'$ then $\bra{\Lambda}\T_{\bar\mu_k}(v)$ is admissible at the point $v'$.
\newline
\newline
Summarising, the recursion \eqref{eq:CHLV2} yields the following recipe for an explicit build up of the eigenstates of the operator $\bB$ with pattern $\Lambda$
\be
\label{eq:generating}
\fbox{
$\displaystyle
\bra{\Lambda^{\bB}}=\bra{0}\prod_{\alpha=1}^L\prod_{k=1}^{\gn-1}\phi^{\gn-k-1}\left(\T_{\bar\mu_k^{\alpha}}(\theta_\alpha+\hbar\,\nu_{k+1}^{\alpha})\right)\,.
$
}
\ee
Here $\bra{0}$ is the lowest weight state (the GT vacuum) of the $\gln$ spin chain, and terms in the product with lower values of $k$ should be left of those with higher values of $k$. We remind the reader that $\phi^{r}$ amounts to the simple replacement of all $T_{ij}$ with $T_{i+r,j+r}$.
\newline
\newline
We should still demonstrate that the constructed states are linearly independent. To this end choose null-twist transfer matrices in \eqref{eq:generating} and use the CBR formula \eqref{cbr} to rewrite them as a sum over products of transfer matricies in anti-symmetric representations. We then take the ASTL \eqref{eq:ASTL} of \eqref{eq:generating}. The leading contribution comes from the term in the CBR expansion with the most number of products\footnote{after using the constraint that the transfer matrix corresponding to the empty diagram $\T_{\es}$ is simply the identity operator}, and it exactly coincides  with the composite raising operator \eqref{compositeraising}. Hence the ASTL of $\bra{\Lambda^{\bB}}$ exists and coincides with $\bra{\Lambda^{\rm GT}}$. So $\bra{\Lambda^{\bB}}$ must be non-zero and moreover all $\bra{\Lambda^{\bB}}$ must be linearly independent for generic enough $w_i$ because $\bra{\Lambda^{\rm GT}}$ are linearly independent. Hence $\bra{\Lambda^{\bB}}$ form a basis (for generic $w_i$) and thus $\bB$ is diagonalisable.
\newline
\newline
One may ask what would happen if $\bar\mu_k^{\alpha}$ in \eqref{actiononadmissible} are chosen to be some arbitrary integer partitions that do not satisfy the branching rules of the GT patterns and hence cannot be interpreted as dual diagonals. Then, if \eqref{actiononadmissible} is non-zero it would be an eigenvector of $\bB$ that is, in general, a linear combination of $\bra{\Lambda^{\bB}}$. Hence the outlined construction \eqref{eq:generating} and generated eigenvectors  $\bra{\Lambda^{\bB}}$ are not unique. However, obvious advantages of the proposed algorithm are that it has clear regular structure and that we can demonstrate that it indeed produces a basis. How one can use this basis is discussed in the next section.

\section{Separation of Variables}
\label{sovsection}\label{sovsection}
In this section we show that the basis \eqref{eq:generating}  leads to separation of variables for the Bethe algebra eigenstates.

 If a basis is generated by action of transfer matrices on some reference state then factorisation of wave functions is immediately obvious \cite{Maillet:2018bim}. One can also use other objects in the Bethe algebra such as Q-operators\footnote{While Q-operators do not belong to the Yangian as an abstract algebra, they do when we descend to representations discussed in this paper. Also note that "other objects" does not mean new conserved charges but rather their repackaging using {\it e.g.} Q-operators instead of transfer matrices.} to reach the same conclusion. However, this is not how the basis \eqref{eq:generating} is constructed currently because lower rank transfer matrices embedded into $\lY(\gln)$ using $\phi$ are typically not elements of the Bethe algebra.
 
  One of the main results to be demonstrated is that we can generate states  \eqref{eq:generating} using auxiliary transfer matricies $\T^{(k)}_{\bar\mu^\alpha_k}$, $k=1,\dots,\gn-1$ who are B{\"a}cklund transforms of the original transfer matrices and who also belong to the Bethe algebra. Namely, we can demonstrate the following equality for any $\bra{\Lambda}\in\lV_{(k)}$
\begin{equation}\label{sov1}
\displaystyle
\bra{\Lambda}\prod_{\alpha=1}^L \phi^{\gn-k-1}\left(\T_{\bar\mu_k^{\alpha}}(\theta_\alpha+\hbar\,\nu_{k+1}^{\alpha})\right)=\bra{\Lambda}\prod_{\alpha=1}^L\T^{(k)}_{\bar{\mu}^\alpha_k}(\theta_\alpha+\hbar\,\nu^\alpha_{k+1})\,.
\end{equation}

We first review the basic properties of the B{\"a}cklund flow in section~\ref{sec:BF} and then focus  on derivation of \eqref{sov1} in section~\ref{sec:ATM}, with some technicalities delegated to appendix \ref{transferaction}. After \eqref{sov1} is established, it is straightforward to use standard Wronskian formulae to get the results about separation of variables announced at the beginning of the paper, as is demonstrated in sections~\ref{sec:WF} and~\ref{sec:CM}.

\subsection{Quantum Eigenvalues, $Q$-system and B{\"a}cklund Flow}
\label{sec:BF}
Given a Young diagram $\xi$ and a group element $g\in\GL(\gn)$ with eigenvalues $z_1,z_2,\dots,z_{\gn}$, its character $\chi_\xi(g)$ in the representation $\xi$ can be obtained from a summation over semi-standard Young tableaux.  A semi-standard Young tableau $\YT$ of shape $\xi$ is obtained by filling up each box in the Young diagram $\xi$ with elements of the set $\{1,2,\dots,\gn\}$ subject to the condition that the numbers weakly decrease in every row and strictly decrease in every column\footnote{Note that our convention is the opposite to the widely used one where the numbers in a tableau strictly increase in each column and weakly increase in each row. The resulting classical character is not sensitive to this difference, however it becomes important for the construction of transfer matrices.}. The character can then be computed as 
\begin{equation}
\chi_\xi(g)=\sum_{\YT}\prod_{(a,s)\subset \xi} z_{\#(a,s)}\,,
\end{equation}
where $\#(a,s)$ denotes the number in position $(a,s)$ of the tableau $\YT$ and the product is over all boxes $(a,s)$ of the diagram $\xi$. 
\newline
\newline
A similar formula exists for transfer matricies \cite{Kuniba:1994na,Tsuboi:1997iq,Tsuboi:1998ne}:
\begin{equation}\label{cbrsoln}
\T_\xi(u)=\sum_{\YT}\prod_{(a,s)\subset \xi} \QEV_{\#(a,s)}(u+\hbar(s-a))\,,
\end{equation}
where the functions $\QEV_j(u),\ j=1,2,\dots,\gn$ are referred to as quantum eigenvalues of the $\lY(\gln)$ monodromy matrix and satisfy 
\begin{equation}
[\QEV_i(u),\QEV_j(v)]=0,\ i,j=1,2,\dots,\gn.
\end{equation}
We will present an explicit construction of them below in terms of another set of quantities, the $Q$-operators \cite{Krichever:1996qd,Tsuboi:2009ud,Bazhanov:2010jq,Kazakov:2010iu,Frassek:2011aa}.

Recall the generating function \eqref{talalaev} for the transfer matricies $\T_{a,1}$: $\det(1-\bT(u)e^{-\hbar \partial_u})=\sum_{a=0}^\gn (-1)^a \T_{a,1}(u)e^{-a\hbar \partial_u}\,.$ It then follows from \eqref{cbrsoln} that we can write
\begin{equation}\label{qeigen}
\det(1-\bT(u)e^{-\hbar \partial_u})=\left(1-\QEV_\gn(u)e^{-\hbar \partial_u} \right)\dots \left(1-\QEV_1(u)e^{-\hbar \partial_u} \right)
\end{equation}
which can easily be seen by expanding the \rhs and comparing coefficients of $e^{-a\hbar \partial_u}$. The $Q$-operators $\Q_i(u)$, $i=1,\dots,\gn$ are annihilated by the above finite-difference operator
\begin{equation}\label{qdefn}
\det(1-\bT(u)e^{-\hbar \partial_u})\Q_i^{[2]}(u)=0,\ i=1,2,\dots,\gn\,.
\end{equation}
The $Q$-operators have been explicitly constructed, by means of various different techniques, in \cite{Bazhanov:2010jq,Kazakov:2010iu,Frassek:2011aa,Bazhanov:1996dr,Derkachov:2003qb,Niccoli:2010sh}. The complete family of Q-operators comprises operators $\Q_I$, $I\subset \{1,2,\dots,\gn\}$ that are related to $\Q_i$ by means of the $QQ$ relations 
\begin{equation}\label{QQ}
\Q_{Iij}\Q_I^{[-2]}=\Q_{Ii}\Q_{Ij}^{[-2]}-\Q_{Ij}\Q_{Ii}^{[-2]}
\end{equation}
supplemented with $\Q_\es(u)=1$. The analytic structure of $Q$-operators for spin chains in arbitrary representation is known \cite{Frassek:2011aa} to have the following form
\begin{equation}\label{qsoln}
\Q_I(u)=N_I\hq_I(u)\prod_{j=1}^{|I|}\Gamma\left[\hat{\nu}_j^{[2(1-|I|)]}(u)\right]\,,\quad \hhq_I(u):=\hq_I \prod_{j\in I}z_j^{\frac u\hbar}\,,
\end{equation}
where $\hat{\nu}_j(u):=\prod_{\alpha=1}^L(u-\theta_\alpha-\hbar\, \hat{\nu}_j^\alpha)$ with $\hat{\nu}_j^\alpha$ being the shifted weights $\hat{\nu}^\alpha_j:=\nu^\alpha_j-j+1$, $\hq_I(u)$ is an operator-valued monic polynomial, and $\hq_{12\ldots\gn}=1$. Finally $N_I$ is normalisation which is well-defined with $N_I=\prod_{j<k}\frac{z_{i_j}-z_{i_k}}{z_{i_j}z_{i_k}}$ for $I=\{i_1,\dots,i_{|I|}\}$ but is not relevant for our discussion, and $\Gamma[F(u)]$ has the property $\Gamma[F(u+\hbar)]=F(u)\Gamma[F(u)]$.

If $I$ is a single index $i$, \eqref{qsoln} becomes
\be\label{gaugetr}
\Q_i(u)=\hhq_i(u)\Gamma\left[{\nu}_1(u)\right]
\ee
which should be considered as a gauge transformation between two ways to parameterise Baxter Q-operators.

By using \eqref{qdefn} together with \eqref{qeigen} it easy to see that a solution for $\QEV_k(u)$ is given by 
\begin{equation}
\label{eq:QEV1}
\QEV_k(u)=\frac{\Q_{\sigma(I_{k-1})}^{[-2]}}{\Q_{\sigma(I_{k-1})}}\frac{\Q_{\sigma(I_{k})}^{[2]}}{\Q_{\sigma(I_{k})}},\quad k=1,\dots,\gn\,,
\end{equation}
where $I_k:=\{1,2,\dots,k\}$, while $\sigma$ denotes some element of the permutation group $\gS_\gn$. Clearly, the quantum eigenvalues $\QEV_k$ are not invariant under choice of $\sigma$ as they are sensitive to the order of terms in the factorisation \eqref{qeigen}. However their (quantum) symmetric combinations, transfer matricies, are invariant under this choice.
\newline
\newline
We will now introduce the notion of the B{\"a}cklund transform. It traces its origins to the solutions of the Hirota bilinear equation on the $\gl(\gn)$ strip \cite{doi:10.1143/JPSJ.45.321,Zabrodin:1996vm,Krichever:1996qd} but we shall define it in more compact terms. Consider the so-called Wronskian solution of the CBR formula \cite{Bazhanov:1996dr,Krichever:1996qd}
\begin{equation}\label{wronsk}
\T_\xi(u)=\frac{\displaystyle\det_{1\leq i,j\leq n}\Q_{\sigma(i)}^{[2\hat{\xi}_{\sigma(j)}]}(u)}{\Q_{\sigma(I_\gn)}(u)}\,,
\end{equation}
where $\hat{\xi}_j=\xi_j-j+1$ are the shifted weights and whose equivalence with \eqref{cbrsoln} follows as a result of the $QQ$-relations. The $(\gn-k)$-th B{\"a}cklund transform of the transfer matrix $\T_\xi(u)$ that shall be denoted as $\T_\xi^{(k)}(u)$ is obtained by restricting the range of the determinant in \eqref{wronsk} to $k$ components:
\begin{equation}\label{wronskian2}
\T_\xi^{(k)}(u)=\frac{\displaystyle\det_{1\leq i,j\leq k}\Q_{\sigma(i)}^{[2\hat{\xi}_{\sigma(j)}]}(u)}{\Q_{\sigma(I_k)}(u)}\,.
\end{equation}
From \eqref{eq:QEV1}, it is easy to deduce that  $\T_\xi^{(k)}$ are expressed in terms of quantum eigenvalues as 
\begin{equation}
\label{cbrsolk}
\T_\xi^{(k)}(u)=\sum_{\YT}\prod_{(a,s)\subset \xi} \QEV_{\#(a,s)}(u+\hbar(s-a))\,,
\end{equation}
where the only difference with \eqref{cbrsoln} is that the tableaux $\YT$ are filled with the numbers $\{1,2,\dots,k\}$, instead of the full set $\{1,2,\dots,\gn\}$. 
\subsection{Action of transfer matrices}
\label{sec:ATM}
We prove \eqref{sov1} in two steps. First, we prove that
\begin{equation}\label{sov2}
\frac{\T_{F^\alpha_k+\bar\mu^\alpha_k}(\theta_\alpha+\hbar\,\nu^\alpha_n)}{\T_{F^\alpha_k}(\theta_\alpha+\hbar\,\nu^\alpha_n)}=\T^{(k)}_{\bar\mu^\alpha_k}(\theta_\alpha+\hbar\,\nu^\alpha_{k+1})\,,
\end{equation}
and then we prove the equality between the \lhs of \eqref{sov2} acting on $\bra{\Lambda}\in\lV_{(k)}$ and the \lhs  \eqref{sov2}. The second step is more technical and we leave it to appendix~\ref{transferaction}, and we also prove in appendix~\ref{invertability} that the ratio of transfer matricies in the \lhs of \eqref{sov2} is well-defined. This subsection deals with \eqref{sov2}.

In our proofs we assume that inhomogeneities assume some generic value (that is we avoid a certain subset of measure zero where the invoked arguments could fail). But since the \lhs of \eqref{sov1} is polynomial in inhomogeneities, the final result should be correct for any $\theta_\alpha$. It is however only useful if \eqref{eq:generating} form a basis for which sake a sufficient condition $\theta_\alpha-\theta_\beta\notin\hbar \ZZ$ for pairwise distinct $\alpha,\beta$ is imposed.

\begin{figure}
\begin{center}
\begin{picture}(170,140)(0,0)
\put(0,0){\includegraphics[width=6cm]{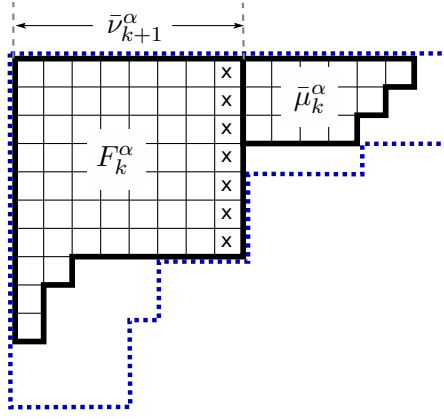}}
\put(36,93){$F^\alpha_k$}
\put(110,116){$\bar{\mu}^\alpha_k$}
\put(40,143){$\bar\nu^\alpha_{k+1}$}
\end{picture}
\end{center}
\caption{\label{fig:Tmu}Gluing of diagrams $F^\alpha_k$ and $\bar{\mu}^\alpha_k$. The dotted line is the boundary of the diagram $\bar\nu^{\alpha}$. Crossed squares depict the column which should be of the same height for $F^{\alpha}_k$ and $\bar\nu^{\alpha}$.}
\end{figure}

In \eqref{sov2}, $\T_{F^\alpha_k+\bar{\mu}^\alpha_k}$ and $\T_{F^\alpha_k}$ are usual $\lY(\gl(\gn))$ transfer matricies and $"+"$ means gluing of Young diagram shapes aligned on top.  Denote by  $\bar\nu^\alpha$ the reduced Young diagram with $\bar\nu^\alpha_j=\nu^\alpha_j-\nu^\alpha_\gn$. Then $F^\alpha_k$ is any Young diagram satisfying the following constraints:  its width (value of the first component $F_{k1}^{\alpha}$) is equal to $\bar\nu^\alpha_{k+1}$, the height of its last column is equal to the height of the $\bar\nu^\alpha_{k+1}$-th column of $\bar\nu^{\alpha}$, and it must be that $F^\alpha_k+\mu^\alpha_k\subset \bar\nu^\alpha$, see Fig~\ref{fig:Tmu}. 

The key feature we need is vanishing of quantum eigenvalues at specific points:
\begin{equation}\label{vanishingeigen}
\QEV_r(\theta_\alpha+\hbar\,\nu_r^\alpha)=0,\ \alpha=1,2,\dots,L,\ r=1,\dots,\gn\,.
\end{equation}
It follows from 
\be
\label{QEVq}
\QEV_r(u)=z_{\sigma(r)}\nu_r(u)\frac{\hq_{\sigma(I_{r-1})}^{[-2]}}{\hq_{\sigma(I_{r-1})}}\frac{\hq_{\sigma(I_{r})}^{[2]}}{\hq_{\sigma(I_{r})}}
\ee 
which can be derived from \eqref{qsoln}, and we assume to avoid situations when the denominator of $\frac{\hq_{\sigma(I_{r-1})}^{[-2]}}{\hq_{\sigma(I_{r-1})}}\frac{\hq_{\sigma(I_{r})}^{[2]}}{\hq_{\sigma(I_{r})}}$ has a pole at $\theta_\alpha+\hbar\,\nu_r^\alpha$.

Consider $\T_\xi(\theta_\alpha+\hbar\,\nu^\alpha_\gn)$ -- the transfer matrix in the representation $\xi$ evaluated at the point $\theta_\alpha+\hbar\,\nu^\alpha_\gn$, and consider its expansion in quantum eigenvalues \eqref{cbrsoln}. For this special point, only a limited subset of tableaux $\YT$ contribute to this expansion. Indeed, let $\YT$ be a tableau that provides a non-zero contribution to the sum. Then it cannot contain $\gn$ at position $a=1,s=1$ because $\QEV_{\gn}(\theta_{\alpha}+\hbar\,\nu_\gn^{\alpha})=0$. But since the numbers in a tableau should weakly decrease to the right and strictly decrease down, $\YT$ cannot contain $\gn$ at all. This tableau cannot also contain $\gn-1$ at position $a=1,s=1+\bar\nu^{\alpha}_{\gn-1}$, due to \eqref{vanishingeigen} for $r=\gn-1$. Then any boxes to the right of the column $s=\bar\nu^{\alpha}_{\gn-1}$ cannot contain $\gn-1$. By repeating the argument we get that boxes of $\YT$ to the right of the column $s=\bar\nu^{\alpha}_{k+1}$ can be populated at most by the indices $1,2,\ldots,k$.

Now we turn to the case when $\xi=F^\alpha_k+\bar{\mu}^\alpha_k$. Let $R$ be the maximal number for which $\bar\nu_{R}^{\alpha}=\bar\nu_{k+1}^{\alpha}$, and $r+1$ be the minimal number for which $\bar\nu_{r+1}^{\alpha}=\bar\nu_{k+1}^{\alpha}$. Then we observe two features. Firstly, entries in the $\bar{\mu}^{\alpha}_k$ part of the tableau $\YT$ can be only populated by indices $1,2,\ldots,r$. Secondly, the height of the last column of $F^\alpha_k$ (denoted by crosses in Fig~\ref{fig:Tmu}) is $R$ and, since $\bar\nu_{R+1}^{\alpha}$ is strictly smaller than $\bar\nu_R^{\alpha}$, this last column can be only populated by indices $1,2,\ldots, R$. Hence it is fixed uniquely. Note that an immediate corollary of this discussion is that $\T_\xi(\theta_\alpha+\hbar\,\nu^\alpha_\gn)=0$ if $\xi$ is any shape not contained in $\bar\nu^\alpha$, in contrast to the fact that the transfer matrix is invertible otherwise as is shown in appendix \ref{invertability}. 

Because for any non-vanishing $\YT$ the last column of the $F^\alpha_k$ part is fixed uniquely, values in other boxes of the $F^\alpha_k$ part do not affect possible values in the boxes of the $\bar{\mu}^{\alpha}_k$ part and vice versa, and so the sum \eqref{cbrsoln} factorises:
\be
\T_{F^\alpha_k+\bar{\mu}^\alpha_k}(\theta_\alpha+\hbar\,\nu^\alpha_\gn)&=&\left(\sum_{\YT_{F}}\prod_{(a,s)\subset F^\alpha_k} \QEV_{\#(a,s)}(\theta_\alpha+\hbar\,\nu^\alpha_\gn+\hbar(s-a))\right)
\nonumber\\
&\times&\left(\sum_{\YT_{\mu}}\prod_{(a,s)\subset \bar{\mu}^\alpha_k} \QEV_{\#(a,s)}(\theta_\alpha+\hbar\,\nu^\alpha_{k+1}+\hbar(s-a))\right)\,.
\ee
The first factor obviously evaluates to $\T_{F^\alpha_k}(\theta_\alpha+\hbar\,\nu^\alpha_\gn)$. For the second one, recall that the possible entries in the tableaux $\YT_{\bar\mu}$ are constrained to be from the set $\{1,2,\ldots,r\}$, but then this term is precisely $\T_{\bar\mu_k^{\alpha}}^{(r)}(\theta_\alpha+\hbar\,\nu^\alpha_{k+1})$, {\it cf.} \eqref{cbrsolk}. By using the same arguments as we invoked after \eqref{vanishingeigen} we show that all $\T_{\bar\mu_k^{\alpha}}^{(k)}(\theta_\alpha+\hbar\,\nu^\alpha_{k+1})$ for $R-1\geq k\geq r$ are in fact equal to one another and hence \eqref{sov2} indeed holds.

We supplement this conclusion with the result of appendix \ref{transferaction} and conclude the remarkable equality \eqref{sov1}. An immediate consequence of \eqref{sov1} is that the basis \eqref{eq:generating} can now be constructed as 
\begin{equation}\label{sov4}
\fbox{
$
\displaystyle
\bra{\Lambda^{\bB}}=\bra{0}\prod_{\alpha=1}^L \prod_{k=1}^{\gn-1}\T^{(k)}_{\bar{\mu}^\alpha_k}(\theta_\alpha+\hbar\,\nu^\alpha_{k+1})
$
}\,.
\end{equation}
We are now one step away from writing concise expressions for wave functions in the SoV basis which is our next goal.

\subsection{Wave functions \& separated variables}
\label{sec:WF}
Expressing the basis \eqref{sov4} using the Wronskian solution \eqref{wronskian2} gives
\begin{equation}\label{sov3}
\bra{\Lambda^{\bf B}}=\bra{0}\prod_{\alpha=1}^L\prod_{k=1}^{\gn-1}\frac{\displaystyle\det_{1\leq i,j\leq k}\Q_{\sigma(i)}^{[2\hat{\bar\mu}_j]}(\theta_\alpha+\hbar\,\nu^\alpha_{k+1})}{\displaystyle \Q_{\sigma(I_k)}(\theta_\alpha+\hbar\,\nu_{k+1}^\alpha)}\,.
\end{equation}
It is convenient to introduce a new reference vector $\bra{\Omega_\sigma}:=\bra{0}\prod\limits_{\alpha=1}^L\prod\limits_{k=1}^{\gn-1}\left( \Q_{\sigma(I_k)}(\theta_\alpha+\hbar\,\nu_{k+1}^\alpha)\right)^{-1} $ for which
\begin{equation}\label{normbasis}
\bra{\Lambda^{\bf B}}=\bra{\Omega_\sigma}\prod_{\alpha=1}^L\prod_{k=1}^{\gn-1}\displaystyle\det_{1\leq i,j\leq k}\Q_{\sigma(i)}(x^\alpha_{kj})\,,
\end{equation}
where we have used that $\svx^\alpha_{kj}=\theta_\alpha+\hbar(\mu^\alpha_{kj}-j+1)$, see \eqref{eq:variablesx}. The Gamma-function contribution to the Q-operators \eqref{gaugetr} nicely factorises from the determinants and we accordingly introduce $\brax$ as rescaled basis vectors $\bra{\Lambda^{\bf B}}$:
\be\label{normbasis2}
\brax:=\prod_{\alpha=1}^L\prod_{k=1}^{\gn-1}\frac 1{\Gamma\left[{\nu}_1(x_{kj}^{\alpha})\right]}\bra{\Lambda^{\bf B}}=\bra{\Omega_\sigma}\prod_{\alpha=1}^L\prod_{k=1}^{\gn-1}\displaystyle\det_{1\leq i,j\leq k}\hhq_{\sigma(i)}(x^\alpha_{kj})\,.
\ee
Let us choose the normalisation $\braket{\Omega_\sigma|\Psi}=1$ for all the Bethe algebra eigenvectors $\ket{\Psi}$. Then their wave functions $\Psi(\svx)$ in the constructed basis are 
\begin{equation}
\label{eq:wav}
\fbox{
$
\displaystyle
\Psi(\svx)=\braket{\svx|\Psi}=\prod_{\alpha=1}^L\prod_{k=1}^{\gn-1}\displaystyle\det_{1\leq i,j\leq k}\hat q_{\sigma(i)}(x^\alpha_{kj})
$
}\,,
\end{equation}
where $\hat q_i(u)$ is the eigenvalue of $\hhq_i(u)$ on the state $\ket{\Psi}$.

With the last formula we achieved our goal of wave function factorisation, and its explicit form justifies why the operators $\svX_{kj}^{\alpha}$ --  zeros of $\bB(u)$ whose eigenvalues on  $\brax$ are $x_{kj}^{\alpha}$ should be considered as separated variables. By choosing $\sigma$ to be the identity permutation we immediately obtain \eqref{wavefn}.
\newline
\newline
Define $\ket{\Omega}$ by the property $\braket{\svx|\Omega}=1$ for all $\brax$. Then \eqref{eq:wav} implies that all $\ket{\Psi}$ can be constructed as 
\begin{equation}
\ket{\Psi}=\prod_{\alpha=1}^L\prod_{k=1}^{\gn-1}\displaystyle\det_{1\leq i,j\leq k}\hat q_{\sigma(i)}(\svX^\alpha_{kj})\ket{\Omega}\,.
\end{equation}
We note that $\ket{\Omega}$ is not itself an eigenvector of the Bethe algebra. In some situations it could be beneficial to select a certain Bethe eigenstate $\ket{0}$ as a reference and build excitations as
\be
\label{ratio}
\ket{\Psi}=\frac{\prod\limits_{\alpha=1}^L\prod\limits_{k=1}^{\gn-1}\displaystyle\det_{1\leq i,j\leq k}\hat q_{\sigma(i)}(\svX^\alpha_{kj})}{\prod\limits_{\alpha=1}^L\prod\limits_{k=1}^{\gn-1}\displaystyle\det_{1\leq i,j\leq k}\hat q_{\sigma(i)}^{(0)}(\svX^\alpha_{kj})}\ket{0}\,,
\ee
where $\hat q_{\sigma(i)}^{(0)}$ is the eigenvalue of $\hhq_{\sigma(i)}$ on $\ket{0}$. The most natural candidate for $\ket{0}$  is one of the ferromagnetic vacua of the spin chain. It is distinguished by the property  $q_{\sigma(12\ldots k)}^{(0)}=1$, $k=1,\ldots,\gn$. In the reference frame where the twist is diagonal it is the highest-weight vector with respect to an appropriate choice of the Borel subalgebra:
\begin{equation}
T_{ij}(u)\ket{0}=0,\quad \sigma^{-1}(i)>\sigma^{-1}(j), \quad T_{jj}(u)\ket{0}=\nu_{\sigma^{-1}(j)}(u) \ket{0}\,,
\end{equation}
and it should be rotated to the modified companion twist frame which we are using in this paper. 

The most drastic simplification of \eqref{ratio} happens  when we consider spin chains in symmetric powers of the fundamental representation. In this case $\nu_j^{\alpha}=0$ for $j>1$ and so, by analysis of section~\ref{sec:ATM}, we can replace $\T^{(k)}_{\bar{\mu}^\alpha_k}$ with $\T^{(1)}_{\bar{\mu}^\alpha_k}$ in  \eqref{sov4}. In particular, $\bar{\mu}^\alpha_k$ consists of a single row. Consequently, \eqref{ratio} becomes
\be
\ket{\Psi}=\frac{\prod\limits_{\alpha=1}^L\prod\limits_{k=1}^{\gn-1}\hat q_{\sigma(1)}(\svX^\alpha_{k1})}{\prod\limits_{\alpha=1}^L\prod\limits_{k=1}^{\gn-1}\displaystyle\hat q_{\sigma(1)}^{(0)}(\svX^\alpha_{kj})}\ket{0}=\prod\limits_{\alpha=1}^L\prod\limits_{k=1}^{\gn-1} q_{\sigma(1)}(\svX^\alpha_{k1})\ket{0}\propto \prod_{r} \bB(u_r)\ket{0}\,,
\ee
where $u_r$ are zeros of $q_{\sigma(1)}$ (the so-called momentum-carrying Bethe roots). We see that, in this special case, $\prod\limits_{r} \bB(u_r)$ acting on the ferromagnetic vacuum creates all the Bethe states. This result was conjectured based on numerical evidence and analytical tests for low numbers of magnons in \cite{Gromov:2016itr} and then proven for $\gl(3)$ \cite{Liashyk:2018qfc} and $\gl(\gn)$ cases \cite{Ryan:2018fyo}.
\newline
\newline
Finally, we make a few comments about the Bethe equations. To simplify our exposition, we will consider all spin chain sites to have the same representation, that is $\nu^\alpha=\nu$ for all $\alpha=1,\dots,L$. In this case it is convenient to introduce the polynomial $Q_\theta(u)=\prod_{\alpha=1}^L(u-\theta_\alpha)$. We also normalise the twist matrix to $\det G=1$.

Originally, the Bethe equations for spin chains in arbitrary representation were written down in \cite{Kulish:1983rd}. These were the equations on zeros of $q_{\sigma(12\ldots)}(u)$ (nested Bethe roots). Instead of such type of Bethe equations, one can write polynomial conditions that should be obeyed by (twisted) polynomials $\hat q_i$. As a consequence of \eqref{QQ} and $Q_{\es}=1$ one derives $\det\limits_{1\leq i,j\leq \gn} Q_i(u-\hbar(j-1))=Q_{12\ldots\gn}$. Then the requirement that $q_{12\ldots\gn}=1$ in \eqref{qsoln} provides a quantisation condition on possible values of $\hat q_i$:
\be
\label{quantcond}
\det\limits_{1\leq i,j\leq \gn}\hat q_i(u-\hbar\,(j-1))\propto\prod_{j=2}^\gn \prod^{\nu_1}_{k=\nu_j+1}Q_\theta(u-\hbar(k+\gn-j))\,,
\ee
where $\propto$ means equality up to a constant multiplication. This quantisation condition is the same as the demand that the Wronskian solution \eqref{wronsk} for transfer matrices $\T_{\xi}$ yields identity if we take $\xi$ to be the empty Young diagram.

There exists also a dual description, in terms of Q-functions $Q^I$ defined by $Q^I:=\varepsilon^{\bar{I}I}Q_{\bar{I}}$, where $\varepsilon$ is the Levi-Civita symbol in $\gn$ dimensions and $\bar{I}$ means the complimentary set to $I$ (no summation over $\bar{I}$ is performed). Again, we can exploit \eqref{QQ} to conclude that $\det\limits_{1\leq i,j\leq\gn}Q^i(u-\hbar(j-1))=\prod\limits_{k=1}^{\gn-1}Q_{12\ldots\gn}(u-\hbar(k-1))$ which, in terms of $\hat q^i:=\varepsilon^{\bar{i}i}\hat q_{\bar{i}}$ becomes
\begin{equation}
\label{hodgequantcond}
\det_{1\leq i,j\leq\gn}\hat q^i(u+\hbar(j-1))\propto\prod_{j=1}^{\gn-1}\prod^{\nu_j}_{k=\nu_\gn+1}Q_\theta(u+\hbar(j-k))\,.
\end{equation}
Note that fixing either $q_i$ or $q^i$ would be sufficient to compute any element of the Bethe algebra.

As was discussed in the introduction, the Bethe algebra is proven to be maximal by existence of the SoV basis. Maximality implies that the above quantisation conditions should have at least as many solutions as the dimension of the Hilbert space, this type of argument can be a powerful tool towards a proof of completeness of Bethe equations, see {\it e.g.} \cite{Niccoli:2009jq,Niccoli:2011nj}. In the case of a spin chain in the defining representation, $\nu=(1,0,\ldots,0)$, the condition \eqref{hodgequantcond} reads $\det\limits_{1\leq i,j\leq \gn}\hat q^i(u+\hbar\,(j-1))\propto Q_\theta(u)$. It  contains only the physical solutions for arbitrary values of inhomogeneities \cite{2013arXiv1303.1578M} and hence can be used alone to fully characterise the spectrum of the model. Similarly, for the conjugate representation $\nu=(1,1,\ldots,1,0)$, the condition \eqref{quantcond} reads $\det\limits_{1\leq i,j\leq \gn}\hat q_i(u-\hbar\,(j-1))\propto Q_\theta(u-\hbar)$ and also is enough to characterise the spectrum.

For more complicated representations than the mentioned two, there are more solutions to \eqref{quantcond} or \eqref{hodgequantcond} than the dimension of the Hilbert space. We should then impose extra restrictions. This can be done by the requirement that $\wT_{\xi}(u)$ should be polynomials in $u$ for any $\xi$ and that $\hq_I(u)$ computed from $\hq_i(u)$ via \eqref{qsoln} and \eqref{QQ} are also polynomials in $u$ for any $I$.  By generalising the ideas of \cite{Marboe:2016yyn} it is possible to repackage these requirements in a structurally simple manner that allows one simple explicit counting of the physical solutions of \eqref{quantcond} and to confirm that their number coincides with the dimension of the Hilbert space. This result will be presented in \cite{LRV}.

\subsection{Conjugate momenta}
\label{sec:CM}
This paper, and also \cite{Ryan:2018fyo}, realises to a large extent Sklyanin's SoV program for compact rational $\gl(n)$ spin chains. Indeed, the operators $X_{kj}^{\alpha}$ are naturally a quantisation of zeros $x_\sigma$ of the classical $B(u)$, and wave functions in the proposed SoV basis are products of determinants of Baxter Q-functions who solve \eqref{qdefn} -- a quantisation of \eqref{eq:HJ}. 

To accomplish the program, we should also quantise $A(u)$ to get the conjugate momenta $P^{\alpha}_{kj}$ and then identify the spin chain with a representation of the algebra generated by $P^{\alpha}_{kj}$ and $\svX^\alpha_{kj}$. Quantisation of $A(u)$ was formally suggested in \cite{Sklyanin:1992sm,2001math.ph...9013S}, however the procedure proposed there becomes singular when explicitly applied to highest-weight spin chains, see for example the discussion in \cite{Maillet:2018bim}. Here we shall introduce conjugate momenta by different means and it would be interesting to explore whether our proposal matches a regularised way to quantise $A(u)$.

The canonically conjugate momenta 
$P^{\pm\alpha}_{kj}$ associated to the separated coordinates $\svX^\alpha_{kj}$  satisfy the commutation relation 
\begin{equation}
[P^{\pm\alpha}_{kj},\svX^\beta_{k'j'}]=\pm \hbar\, \delta^{\alpha\beta}\delta_{kk'}\delta_{jj'} P^{\pm\alpha}_{kj}.
\end{equation}
We propose their following realisation
\begin{equation}\label{raiselower}
P^{\pm\alpha}_{kj}=c^{\pm\alpha}_{kj} :\frac{\displaystyle\det_{1\leq i,l\leq k} \Q_{\sigma(i)}(\svX^\alpha_{kl}\pm\hbar\delta_{jl})}{\displaystyle\det_{1\leq i,l\leq k} \Q_{\sigma(i)}(\svX^\alpha_{kl})}:\,,
\end{equation}
where $c^{\pm\alpha}_{kj}$ is some simple function of the separated variables to be fixed in a moment. We use a normal ordering prescription $:\ :$ where $\svX$'s are placed to the left of all the coefficients of Baxter $Q$-operators. 
To see that the prescription \eqref{raiselower} works, we utilise \eqref{sov3} and act on $\brax$ with $P^{\pm\alpha}$ as defined above. By using that $\brax \svX^\alpha_{kj}=\svx^\alpha_{kj}\brax$, we immediately obtain (up to normalisation) the state where $\mu^\alpha_{kj}$ has been replaced with $\mu^\alpha_{kj}\pm 1$. In particular the action of $P^{\pm\alpha}_{kj}$ on $\brax$ is well-defined. 
\newline
\newline
The coefficient $c^{\pm\alpha}_{kj}$  in \eqref{raiselower} is required in order to respect the branching rules of GT patterns. Namelly, we have the constraints $\mu^\alpha_{k-1,j}\geq\mu^\alpha_{kj}\geq \mu^\alpha_{k,j+1}$ and $\mu^\alpha_{k,j-1}\geq\mu^\alpha_{kj}\geq \mu^\alpha_{k+1,j}$ on a given GT pattern $\Lambda^\alpha$  and so $P^{+\alpha}_{kj}$ should vanish when we act on a state with $\mu^\alpha_{kj}=\mu^\alpha_{k,j-1}$ or $\mu^\alpha_{kj}=\mu^\alpha_{k-1,j}$, and similarly for $P^{-\alpha}_{kj}$. Using the fact that $\mu^\alpha_{kj}$ is related to $\svx^\alpha_{kj}$ as $\svx^\alpha_{kj}=\theta_\alpha+\hbar(\mu^\alpha_{kj}-j+1)$ we see that we should take \begin{equation}
c^{+\alpha}_{kj}=(\svX^\alpha_{k-1,j}-\svX^\alpha_{kj})(\svX^\alpha_{k,j-1}-\svX^\alpha_{kj}-\hbar)
\end{equation}
and similarly
\begin{equation}
c^{-\alpha}_{kj}=(\svX^\alpha_{kj}-\svX^\alpha_{k+1,j})(\svX^\alpha_{kj}-\svX^\alpha_{k,j+1}-\hbar)\,.
\end{equation}
The separated variables $\svX^\alpha_{kj}$ are defined for indices in the range $1\leq k\leq \gn-1$ and $1\leq j\leq k$, but $c^{\pm\alpha}_{kj}$ can contain factors with $\svX^\alpha_{kj}$ outside of this range. In order to get around this we define operators $\svX^\alpha_{j,j+1}$, $j=0,\dots,\gn-1$ to be scalar multiples of the identity operator with eigenvalue $\theta_\alpha+\hbar(\nu^\alpha_{j+1}-j)$. Furthermore, if $c^{\pm\alpha}_{kj}$ should contain a factor with $\svX^\alpha_{kj}$ outside of this newly established set of operators, we simply declare that factor to be absent. 
\section{Outlook}
Now that we have access to the wave functions of the Bethe algebra the next obvious step is to use the obtained results to compute scalar products and form factors of various operators. Scalar products in the SoV approach have previously been considered for the $\gl(2)$ case in \cite{Kazama:2013rya,Kitanine:2015jna}. These results were generalised in \cite{Gromov:2019wmz} for the defining representation of $\gl(3)$ by introducing a second set of separated variables $\svY^\alpha_{kj}$ as operatorial roots of a $\bC$-operator whose right eigenstates factorise the left eigenstates of the Bethe algebra. This was then used to compute the scalar product between two Bethe states, in agreement with the functional orthogonality approach developed in \cite{Cavaglia:2018lxi,Cavaglia:2019pow}. Generalisation of this interplay between operatorial and functional scalar products was then subsequently extended to $\gla(\gn)$ spin chains in \cite{Gromov:2020fwh}.
\newline
\newline
The focus of this work has been on compact spin chains. An open question is the generalisation of the discussed techniques to the case of non-compact and supersymmetric spin chains, such as those with $\mathfrak{su}(p,q|m)$ symmetry necessary for AdS/CFT applications. The computation of scalar products and form-factors in the non-compact case was considered in \cite{Cavaglia:2019pow} based on the functional formalism, and it was related to an operatorial constriction of states in \cite{Gromov:2020fwh} for a certain class of non-compact highest-weight representations, similar to what was done here, and it would be interesting to extend this procedure to all highest-weight representations. An SoV basis was constructed in \cite{Maillet:2019ayx} for the case of the defining representation of $\gl(\gm|\gn)$ super spin chains and the Hubbard model, and it would be interesting to attempt relating the constructed basis to the $\bB$-type operator constructed in \cite{Gromov:2018cvh}, as well as generalise findings beyond the fundamental representation, as it was done here in the bosonic setting. 

One should also generalise the discussed techniques to models based on the principal series representations of $\gln$. The SoV framework for models with principal series representations of $\gl(2)$ has been carried out in \cite{Derkachov:2001yn,Derkachov:2002tf}, with some initial progress being made for the $\gl(3)$ case in \cite{Derkachov:2018ewi}. A feature of the principal series setting is that, in contrast to the compact case, it is not necessary to introduce a boundary twist in order for the $\bB$-operator to be diagonalisable, and hence such a twist is not usually employed. However, doing so may be beneficial as the $\bB$-operator can still be related to the Gelfand-Tsetlin subalgebra with the use of the companion twist. Study of the Gelfand-Tsetlin subalgebra in the principal series setting was carried out in \cite{Valinevich:2016cwq,ValinevichGT}. The SoV framework in the principal series setting of $\gl(2)$ was recently utilised in \cite{Derkachov:2018rot} for the computation of Basso-Dixon correlators in two-dimensional fishnet CFT \cite{Gurdogan:2015csr,Kazakov:2018qez} and a set of separated variables for the case of $\mathfrak{so}(1,5)$ spin chains were constructed in \cite{Derkachov:2019tzo} which are related to the computations of \cite{Basso:2019xay}.
\newline
\newline
Finally, it would be interesting to extend our results to other quantum integrable models. In particular, an SoV basis for the case of $U_q(\widehat{\mathfrak{sl}}(n))$ was constructed in \cite{Maillet:2018rto} and it would be interesting to check if it diagonalises the $\bB$-operator proposed in \cite{2001math.ph...9013S}.
\paragraph{Acknowledgements} We are grateful to D.Chernyak, F.Levkovich-Maslyuk,  N.Gromov and especially S. Leurent  for useful discussions. The work of P.R. is partly supported by a Nordita Visiting Ph.D Fellowship and by SFI and the Royal Society grant UF160578. The work of D.V. is supported by the Knut and Alice Wallenberg Foundation under grant Dnr KAW 2015.0083.
\appendix

\section{Invertability of transfer matricies}\label{invertability}
Here we prove that $\T_\xi(\theta_\alpha+\hbar\,\nu_\gn^\alpha)$ is invertible when $\xi\subset \bar{\nu}^\alpha$, where $\bar{\nu}^\alpha$ denotes the reduced Young diagram $\bar\nu_j^\alpha=\nu_j^\alpha-\nu_\gn^\alpha$, $j=1,\dots,\gn$. We will see below that provided inhomogeneities are largely separated, that is $|\theta_\alpha-\theta_\beta|\gg 1$ for $\alpha\neq \beta$ then the transfer matricies effectively become equal to those of $L=1$. Hence, we start by considering this case. Any given transfer matrix $\T_\xi(u)$ is a polynomial in $\theta_\alpha$ and the entries of the twist matrix $G$. Hence if we can prove the claim for a specific value of the twist then it must be true generically, {\it i.e}. away from some measure zero subset. To this end, let us make use of the fact that transfer matricies are central for $L=1$ when $G=1$ where the computation simplifies. In what follows we will omit the $\alpha$-index.
\newline
\newline
A convenient tool to prove the claim is the quantum eigenvalues introduced in section \ref{sovsection}. By acting on the highest-weight state it is easy to see that $\Lambda_j(u)=(u-\theta-\hbar\, \nu_j)$. 
The transfer matrix $\T_{a,1}(u)$ can be written as a sum over quantum semi-standard Young tableaux of the form 
\begin{equation}
\ytableausetup{centertableaux}
\begin{ytableau}
i_a \\
\none[\vdots] \\
i_2 \\ 
i_1 \\
\end{ytableau}
\end{equation}
subject to the constraint $i_1<i_2<\dots <i_a$. By using the recipe to assign products of quantum eigenvalues to a tableau  we associate the factor $\prod_{k=1}^a(u-\theta-\hbar(\nu_{i_k}+a-k))$ to the above tableau. Let us now evaluate this factor at $\theta+\hbar\,\nu_\gn$. We obtain 
\begin{equation}
(-\hbar)^a(\bar\nu_{i_a})(\bar\nu_{a-1}+1)\dots(\bar\nu_{i_1}+a-1)\,.
\end{equation}
Since $\bar\nu_j\geq 0$ for all $j=1,\dots,\gn$ it follows that the above expression is non-negative. Note that if some weight $\nu_k=\nu_\gn$, it forces $\bar\nu_k=\bar\nu_{k+1}=\dots=\bar\nu_\gn=0$ and hence the indices $k,k+1,\dots,\gn$ cannot appear in the tableau as they provide vanishing contributions. Hence, in order to have a non-vanishing term we must at least have $\bar\nu_a\geq 1$ and hence $\bar\nu_1\geq \bar\nu_2\geq \dots \geq \bar\nu_a\geq 1$. Hence, $\T_{a,1}(\theta+\hbar\,\nu_\gn)$ is non-zero if 
\begin{equation}
(1^a)\subset \bar{\nu}\,.
\end{equation}
Now we consider an arbitrary Young diagram $\xi$. $\T_\xi(\theta+\hbar\,\nu_\gn)$ can be written as a sum over Young tableaux as before, and we will consider the factors of quantum eigenvalues associated to each column separately. The admissible indices such that a given column is non-vanishing directly effects what indices can appear in the columns to the right. Indeed, we already know the first column will always be non-negative, and we will get a non-zero contribution if 
\begin{equation}
(1^{\xi^{\rm T}_1})\subset \bar{\nu}\,.
\end{equation}
Now we go to the second column which gives the contribution 
\begin{equation}
(-\hbar)^{\xi^{\rm T}_2}(\bar\nu_{i_{\xi^{\rm T}_2}}-1)(\bar\nu_{i_{\xi^{\rm T}_2}-1}-2)\dots (\bar\nu_{i_{1}}+\xi^{\rm T}_2-2)\,.
\end{equation}
Since the first column is non-zero, if we put some number $k$ in the top box of the second column we must have that $\bar\nu^\alpha_k> 1$ and hence the second column will be non-zero if 
\begin{equation}
\bar\nu_1\geq \bar\nu_2\geq \dots\geq \bar\nu_{\xi^{\rm T}_2}\geq 2\,.
\end{equation}
Hence, the contribution from the first two columns will be non-zero if 
\begin{equation}
(1^{\xi^{\rm T}_1}1^{\xi^{\rm T}_2})\subset \bar{\nu}\,.
\end{equation}
Continuing in the same way, we find that if $\xi\subset \bar{\nu}$ there will always be a tableau which does not vanish and the signs of the contributions of all non-vanishing tableaux are all the same and equal to the sign of $(-1)^{|\xi|}$, where $|\xi|$ denotes the number of boxes in the Young diagram $\xi$. Hence for $L=1$ $\T_\xi(\theta+\hbar\,\nu_n)$ is non-zero. 
\newline
\newline
Now we consider $L>1$. The transfer matrix $\T_\xi$ is obtained by taking the trace of the fused monodromy matrix $T_\xi(u)$ which itself is a product of fused $R$-matricies $R^{\xi,\nu^\alpha}$
\begin{equation}
\T_\xi(u)=\displaystyle\sum_{i_1,\dots,i_L}R^{\xi,\nu^1}_{i_1 i_2}(u-\theta_1)\otimes \dots \otimes R^{\xi,\nu^L}_{i_L i_1}(u-\theta_L)\,,
\end{equation}
where the sum ranges over $1,2,\dots,{\rm dim }\,\xi$. Since $R^{\xi,\nu^\beta}(u)\sim u^{|\xi|}$ at large $u$, with $|\xi|$ denoting the number of boxes in the Young diagram $\xi$, we can consider $\T_\xi(\theta_\alpha+\hbar\,\nu^\alpha_n)$ in the limit $|\theta_\beta-\theta_\alpha|\gg 1$ for all $\beta\neq \alpha$. In this limit $\T_\xi(\theta_\alpha+\hbar\,\nu^\alpha_n)$ coincides (up to irrelevant normalisation) with the $L=1$ transfer matrix which we know is invertible and so $\T_\xi(\theta_\alpha+\hbar\,\nu_n^\alpha)$ is invertible for generic values of inhomogeneities, completing the proof. 
\section{Action of transfer matricies -- technical details}\label{transferaction}
We need to prove that 
\be
\label{needtoprove}
\bra{\Lambda}\prod_{\alpha=1}^L\displaystyle\frac{\displaystyle \T_{F^\alpha_k+\bar{\mu}^\alpha_k}(\theta_\alpha+\hbar\,\nu^\alpha_n)}{\displaystyle \T_{F^\alpha_k}(\theta_\alpha+\hbar\,\nu^\alpha_n)}=\bra{\Lambda}\prod_{\alpha=1}^L \phi^{\gn-k-1}\left(\T_{\bar\mu_k^{\alpha}}(\theta_\alpha+\hbar\,\nu_{k+1}^{\alpha})\right)\,
\ee
if $\bra{\Lambda}\in\lV_{(k)}$. This result easily follows from the following one which we are going to prove: For a state of the form
\begin{equation}
\bra{\Lambda_I}:=\bra{\Lambda}\prod_{\gamma\in I}\phi^{\gn-k-1}\left(\T_{\bar\mu^\gamma_{k}}(\theta_\gamma+\hbar\,\nu^\gamma_{k+1})\right)\,,
\end{equation}
where $\bra{\Lambda}\in\lV_{(k)}$ and $I$ is a subset of $\{1,\ldots,L\}$, it is true that
\begin{equation}
\bra{\Lambda_I}\phi\left(\T_{\bar\mu^\alpha_{k}}(\theta_\alpha+\hbar\,\nu^\alpha_{k+1})\right)=\bra{\Lambda_I}\frac{\T_{R_{\gn-1}+\dots+R_{k-1}+\bar\mu^\alpha_k}}{\T_{R_{\gn-1}+\dots+R_{k-1}}}
\end{equation}
for $\alpha\notin I$. Here both transfer matrices on the \rhs are evaluated at $\theta_\alpha+\hbar\,\nu^\alpha_\gn$, and $R_{\gn-1}+\dots+R_{k-1}$ is a specific choice of Young diagram $F_k^{\alpha}$ to be made precise below\footnote{Recall that the ratio in the \lhs of \eqref{needtoprove} is invariant under variations of $F_k^{\alpha}$ subject to certain constraints, we are making one particular choice that simplifies computations.}.
\newline
\newline
We will need two technical results. First, let us note that quantum minors satisfy the following commutativity property \cite{molev2007yangians}. If $\lA$ and $\lB$ are subsets of $\{1,2,\dots,\gn\}$ then
\begin{equation}\label{commutativity}
[T\left[^\lA_\lB\right](u),T_{ab}(v)]=0
\end{equation}
for all $a\in\lA$ and $b\in\lB$. Next, suppose $\bra{\Lambda}$ of $\GT_1,\dots,\GT_r$ for some $r$, for which the dual diagonal $\mu^\alpha_r$ takes its minimal allowed value $\mu^\alpha_{rj}=\nu^\alpha_{r+1}$, $j=1,\dots,r$ and $\mu^\alpha_{r+1}$ takes its maximal allowed value given the previous constraint $\mu^\alpha_{r+1,j}=\nu^\alpha_{r+1}$, $j=1,\dots,r+1$. Then we have
\begin{equation}\label{shortening}
\bra{\Lambda}T_{j,\gn-r}(\theta_\alpha+\hbar\,\mu^\alpha_{\gn-r+1,1})=0,\quad j=\gn-r-1,\dots,\gn
\end{equation}
which is simply the statement that the dual diagonal $\mu^\alpha_{r+1}$ cannot be excited further without changing $\mu^\alpha_r$ and that $\mu^\alpha_r$ cannot be lowered without first lowering $\mu^\alpha_{r+1}$. The proof of this is very similar to that of the statements (3.36-3.38) in \cite{Ryan:2018fyo} adapted to this more general setting and so we do not repeat it here. The motivation for this statement is that when we act with transfer matricies $\T_{R_{\gn-1}+\dots+R_{k-1}+\bar\mu^\alpha_k}$ the action on $\bra{\Lambda_I}$ will factorise, and each $\T_{R_j}$ factor will act as a raising operator exciting a dual diagonal to its maximal where it is equal to the next dual diagonal, allowing us to use the previous result.
\newline
\newline
Let $\bar{\nu}^\alpha$ denote the reduced Young diagram $\bar{\nu}^\alpha_j=\nu^\alpha_j-\nu^\alpha_\gn$, $j=1,\dots,\gn-1$. 
$\bar\nu^\alpha$ splits into the rectangular regions $R_j$, $j=1,\dots,\gn-1$, where the width of $R_j$ is $\bar\nu^\alpha_j-\bar\nu^\alpha_{j+1}$ and its height is $j$. By $R_{\gn-1}+\dots+R_{k-1}$ we denote the subdiagram of $\bar\nu^\alpha$ comprising the first $\bar\nu_{k-1}^{\alpha}$ columns of $\bar\nu^\alpha$. Note that the state $\bra{\Lambda_I}$ is an admissible vector at point $\theta_\alpha+\hbar\,\nu^\alpha_\gn$ and so the action of $\T_{R_{\gn-1}+\dots+\bar\mu^\alpha_k}(\theta_\alpha+\hbar\,\nu^\alpha_\gn)$ with the MCT \eqref{modcomp} coincides with that of the null twist, {\it cf.} page~\pageref{pos:admissible}.
\newline
\newline
For simplicity of exposition, we will assume that all weights $\nu^\alpha_j$ are distinct, and will comment later on what happens when they are not. For all weights being distinct, the region $R_j$ has non-vanishing width and furthermore we have the following factorisation 
\begin{equation}\label{factorout}
\T_{R_{\gn-1}+\dots+\bar\mu^\alpha_k}(u)=\T_{R_{\gn-1}}(u)\T_{R_{\gn-2}+\dots+\bar\mu^\alpha_{k}}(u+\hbar\,\bar\nu^\alpha_{\gn-1})\,.
\end{equation}
To see this we utilise the CBR formula \eqref{cbr} which says that for some Young diagram $\xi$ one has
\begin{equation}
\T_\xi(u)=\sum_{\sigma\in S_{\gn}} \T_{\xi_1^{\rm T}+\sigma(1)-1,1}(u+\hbar(\sigma(1)-1))\times \dots\,.
\end{equation}
When we use the null twist, all $\xi$ are constrained to have height at most $\gn-1$, and for the case of interest to us we have $\xi^{\rm T}_1=\gn-1$. In the above sum, if for some permutation $\sigma$ we have $\sigma(1)\neq 1$ then $\sigma(1)>1$ and so the sum contains a transfer matrix of height greater than $\gn-1$ and so must vanish. Hence, we must have that the transfer matrix factorises into $\T_{\xi_1^{\rm T}}(u)\times \dots$ where $\dots$ refers to the transfer matrix corresponding to the Young diagram obtained from $\xi$ by removing its first column. If the second column also has height $\gn-1$ then it also factors out and so on. Hence \eqref{factorout} follows, where now
\begin{equation}\label{Tfactor}
\T_{R_{n-1}}(u)=\T_{n-1,1}(u)\dots \T_{n-1,1}(u+\hbar(\bar\nu^\alpha_{n-1}-1))\,,
\end{equation}
and so the \rhs \eqref{Tfactor} coincides with the composite raising operator \eqref{compositeraising} for the right-most dual diagonal. Hence, evaluating at $u=\theta_\alpha+\hbar\,\nu^\alpha_\gn$ we see that acting with $\T_{R_{\gn-1}}$ takes us from $\bra{\Lambda_I}$ to the state $\bra{\Lambda'_I}$ with $\mu^\alpha_{\gn-1,j}=\mu^\alpha_{\gn-2,j}=\nu^\alpha_{\gn-1}$, $j=1,\dots,\gn-2$ and $\mu^\alpha_{\gn-1,\gn-1}=\nu^\alpha_{\gn-1}$ which satisfies {\eqref{shortening}.

The action of $\T_{R_{n-2}+\dots+\bar\mu^\alpha_{k}}(u+\hbar\bar\nu^\alpha_{n-1})$ on $\bra{\Lambda'_I}$ is expressed as a sum over tableaux $\sum_{\lA}T\left[^\lA_{\lA+1}\right]$ where $\lA+1$ cannot contain the number $2$ by \eqref{shortening}, and so $\lA$ cannot contain $1$, forbidding us from having transfer matricies of size $\gn-1$ and so the action again factorises into 
\begin{equation}
\bra{\Lambda_I}\T_{R_{\gn-1}}\T_{R_{\gn-2}}\T_{R_{\gn-3}+\dots+\bar\mu^\alpha_{k}}(u+\hbar\bar\nu^\alpha_{\gn-2})\,.
\end{equation}
Hence when the $\T_{R_{\gn-2}}$ factor acts on $\bra{\Lambda}\T_{R_{\gn-1}}$ it will excite the dual diagonals to the configuration where $\mu^\alpha_{\gn-2,j}=\mu^\alpha_{\gn-3,j}=\nu^\alpha_{\gn-2}$, $j=1,\dots,\gn-3$ and $\mu^\alpha_{\gn-2,\gn-2}=\nu^\alpha_{\gn-2}$ and again the results of \eqref{shortening} apply, further limiting the indicies which can populate the tableaux making up the $\T_{R_{\gn-3}+\dots}$ factor. 

The end result is that the action of $\T_{R_{\gn-1}+\dots}$ completely factorises into 
\begin{equation}
\bra{\Lambda_I}\T_{R_{\gn-1}}\T_{R_{\gn-2}}\dots \T_{R_{k-1}} \T_{\bar\mu^\alpha_k}(\theta_\alpha+\hbar\,\nu^\alpha_{k+1})\,,
\end{equation}
where we have omitted the spectral parameters of the $\T_{R_j}$ factors for brevity and $\T_{\bar\mu^\alpha_k}$ should be understood as $\sum_{\lA}T_{\bar\mu^\alpha_k}$ where $\lA$ can only be populated with indices from the set $\{\gn-k,\dots,\gn-1\}$. Then, using \eqref{commutativity} we can move this factor to the left, obtaining 
\begin{equation}
\begin{split}
& \bra{\Lambda_I}\T_{R_{\gn-1}+\dots+\bar\mu^\alpha_k}(\theta_\alpha+\hbar\,\nu^\alpha_\gn) \\ &=\bra{\Lambda_I}\phi^{\gn-k-1}\left(\T_{\bar\mu^\alpha_k}(\theta_\alpha+\hbar\,\nu^\alpha_{k+1})\right)\T_{R_{\gn-1}+\dots+R_{k-1}}(\theta_\alpha+\hbar\,\nu^\alpha_{\gn})\,.
\end{split}
\end{equation}
This completes the proof since invertiblity of the transfer matrix was proven in the previous appendix.
\newline
\newline
Finally, let us briefly discuss the case of coinciding weights. As we have seen above, each factorisation into a rectangular region results in a reduction of the number of indices in the factors which appear to the right of it. If two weights coincide, say $\nu^\alpha_j=\nu^\alpha_{j+1}$ then the rectangle $R_j$ has vanishing width and so does not contribute to the factorisation. One could then expect that at the end the right most factor could contain more than just the indices $\gn-k,\dots,\gn-1$, ruining our conclusion. However, if two weights coincide then $\bra{\Lambda_I}$ will have extra dual diagonals  $\mu_{k+1}^{\alpha}, \mu_{k+2}^{\alpha},\ldots$ whose entries are all equal to $\nu_{k+1}^{\alpha}$. They will extend the range of indices in \eqref{shortening} which annihilate $\bra{\Lambda}$ similar to the case of rectangular representations discussed in \cite{Ryan:2018fyo}, which will further constrain the indices that can appear in the sum over tableaux. Taking this into account we find that the end conclusion is the same. 

\bibliographystyle{utphys}
\bibliography{References}

\providecommand{\href}[2]{#2}\begingroup\raggedright\begin{thebibliography}{10}

\bibitem{10.1007/3-540-15213-X_80}
E.~K. Sklyanin, ``The quantum toda chain,'' in {\em Non-Linear Equations in
  Classical and Quantum Field Theory} (N.~Sanchez, ed.), (Berlin, Heidelberg),
  pp.~196--233, Springer Berlin Heidelberg, 1985.

\bibitem{Sklyanin:1991ss}
E.~K. Sklyanin, ``{Quantum inverse scattering method. Selected topics},''
  \href{http://xxx.lanl.gov/abs/hep-th/9211111}{{\tt hep-th/9211111}}.

\bibitem{2001math.ph...9013S}
F.~A. {Smirnov}, ``{Separation of variables for quantum integrable models
  related to $U_q(\hat{sl}_N) $},'' {\em arXiv e-prints} (Sep, 2001)
  math--ph/0109013, \href{http://xxx.lanl.gov/abs/math-ph/0109013}{{\tt
  math-ph/0109013}}.

\bibitem{Gromov:2016itr}
N.~Gromov, F.~Levkovich-Maslyuk, and G.~Sizov, ``{New Construction of
  Eigenstates and Separation of Variables for SU(N) Quantum Spin Chains},''
  {\em JHEP} {\bf 09} (2017) 111,
  \href{http://xxx.lanl.gov/abs/1610.08032}{{\tt 1610.08032}}.

\bibitem{Maillet:2018bim}
J.~M. Maillet and G.~Niccoli, ``{On quantum separation of variables},'' {\em J.
  Math. Phys.} {\bf 59} (2018), no.~9 091417,
  \href{http://xxx.lanl.gov/abs/1807.11572}{{\tt 1807.11572}}.

\bibitem{Ryan:2018fyo}
P.~Ryan and D.~Volin, ``{Separated variables and wave functions for rational
  gl(N) spin chains in the companion twist frame},'' {\em J. Math. Phys.} {\bf
  60} (2019), no.~3 032701, \href{http://xxx.lanl.gov/abs/1810.10996}{{\tt
  1810.10996}}.

\bibitem{Gromov:2013pga}
N.~Gromov, V.~Kazakov, S.~Leurent, and D.~Volin, ``{Quantum Spectral Curve for
  Planar $\mathcal{N} = 4$ Super-Yang-Mills Theory},'' {\em Phys. Rev. Lett.}
  {\bf 112} (2014), no.~1 011602, \href{http://xxx.lanl.gov/abs/1305.1939}{{\tt
  1305.1939}}.

\bibitem{Gromov:2014caa}
N.~Gromov, V.~Kazakov, S.~Leurent, and D.~Volin, ``{Quantum spectral curve for
  arbitrary state/operator in AdS$_{5}$/CFT$_{4}$},'' {\em JHEP} {\bf 09}
  (2015) 187, \href{http://xxx.lanl.gov/abs/1405.4857}{{\tt 1405.4857}}.

\bibitem{Cavaglia:2018lxi}
A.~Cavaglià, N.~Gromov, and F.~Levkovich-Maslyuk, ``{Quantum spectral curve
  and structure constants in $ \mathcal{N}=4 $ SYM: cusps in the ladder
  limit},'' {\em JHEP} {\bf 10} (2018) 060,
  \href{http://xxx.lanl.gov/abs/1802.04237}{{\tt 1802.04237}}.

\bibitem{Derkachov:2001yn}
S.~E. Derkachov, G.~P. Korchemsky, and A.~N. Manashov, ``{Noncompact Heisenberg
  spin magnets from high-energy QCD: 1. Baxter Q operator and separation of
  variables},'' {\em Nucl. Phys.} {\bf B617} (2001) 375--440,
  \href{http://xxx.lanl.gov/abs/hep-th/0107193}{{\tt hep-th/0107193}}.

\bibitem{Derkachov:2002tf}
S.~E. Derkachov, G.~P. Korchemsky, and A.~N. Manashov, ``{Separation of
  variables for the quantum SL(2,R) spin chain},'' {\em JHEP} {\bf 07} (2003)
  047, \href{http://xxx.lanl.gov/abs/hep-th/0210216}{{\tt hep-th/0210216}}.

\bibitem{Derkachov:2018rot}
S.~Derkachov, V.~Kazakov, and E.~Olivucci, ``{Basso-Dixon Correlators in
  Two-Dimensional Fishnet CFT},'' {\em JHEP} {\bf 04} (2019) 032,
  \href{http://xxx.lanl.gov/abs/1811.10623}{{\tt 1811.10623}}.

\bibitem{Basso:2019xay}
B.~Basso, G.~Ferrando, V.~Kazakov, and D.-l. Zhong, ``{Thermodynamic Bethe
  Ansatz for Fishnet CFT},'' \href{http://xxx.lanl.gov/abs/1911.10213}{{\tt
  1911.10213}}.

\bibitem{Derkachov:2019tzo}
S.~Derkachov and E.~Olivucci, ``{Exactly solvable magnet of conformal spins in
  four dimensions},'' \href{http://xxx.lanl.gov/abs/1912.07588}{{\tt
  1912.07588}}.

\bibitem{Gromov:2019aku}
N.~Gromov and A.~Sever, ``{Derivation of the Holographic Dual of a Planar
  Conformal Field Theory in 4D},'' {\em Phys. Rev. Lett.} {\bf 123} (2019),
  no.~8 081602, \href{http://xxx.lanl.gov/abs/1903.10508}{{\tt 1903.10508}}.

\bibitem{Gromov:2019bsj}
N.~Gromov and A.~Sever, ``{Quantum fishchain in AdS$_{5}$},'' {\em JHEP} {\bf
  10} (2019) 085, \href{http://xxx.lanl.gov/abs/1907.01001}{{\tt 1907.01001}}.

\bibitem{Gromov:2019jfh}
N.~Gromov and A.~Sever, ``{The Holographic Dual of Strongly $\gamma$-deformed
  N=4 SYM Theory: Derivation, Generalization, Integrability and Discrete
  Reparametrization Symmetry},'' \href{http://xxx.lanl.gov/abs/1908.10379}{{\tt
  1908.10379}}.

\bibitem{Gunaydin:2017lhg}
M.~Günaydin and D.~Volin, ``{The complete unitary dual of non-compact Lie
  superalgebra su(p,q|m) via the generalised oscillator formalism, and
  non-compact Young diagrams},'' \href{http://xxx.lanl.gov/abs/1712.01811}{{\tt
  1712.01811}}.

\bibitem{Marboe:2017dmb}
C.~Marboe and D.~Volin, ``{The full spectrum of AdS5/CFT4 I: Representation
  theory and one-loop Q-system},'' {\em J. Phys.} {\bf A51} (2018), no.~16
  165401, \href{http://xxx.lanl.gov/abs/1701.03704}{{\tt 1701.03704}}.

\bibitem{Sklyanin:1992sm}
E.~K. Sklyanin, ``{Separation of variables in the quantum integrable models
  related to the Yangian Y[sl(3)]},'' {\em J. Math. Sci.} {\bf 80} (1996)
  1861--1871, \href{http://xxx.lanl.gov/abs/hep-th/9212076}{{\tt
  hep-th/9212076}}. [Zap. Nauchn. Semin.205,166(1993)].

\bibitem{Maillet:2018czd}
J.~M. Maillet and G.~Niccoli, ``{Complete spectrum of quantum integrable
  lattice models associated to Y(gl(n)) by separation of variables},'' {\em
  SciPost Phys.} {\bf 6} (2019) 071,
  \href{http://xxx.lanl.gov/abs/1810.11885}{{\tt 1810.11885}}.

\bibitem{Maillet:2019nsy}
J.~M. Maillet and G.~Niccoli, ``{On quantum separation of variables beyond
  fundamental representations},''
  \href{http://xxx.lanl.gov/abs/1903.06618}{{\tt 1903.06618}}.

\bibitem{Krichever:1996qd}
I.~Krichever, O.~Lipan, P.~Wiegmann, and A.~Zabrodin, ``{Quantum integrable
  systems and elliptic solutions of classical discrete nonlinear equations},''
  {\em Commun. Math. Phys.} {\bf 188} (1997) 267--304,
  \href{http://xxx.lanl.gov/abs/hep-th/9604080}{{\tt hep-th/9604080}}.

\bibitem{Kazakov:2007fy}
V.~Kazakov, A.~S. Sorin, and A.~Zabrodin, ``{Supersymmetric Bethe ansatz and
  Baxter equations from discrete Hirota dynamics},'' {\em Nucl. Phys.} {\bf
  B790} (2008) 345--413, \href{http://xxx.lanl.gov/abs/hep-th/0703147}{{\tt
  hep-th/0703147}}.

\bibitem{Zabrodin:2007rq}
A.~Zabrodin, ``{Backlund transformations for difference Hirota equation and
  supersymmetric Bethe ansatz},'' \href{http://xxx.lanl.gov/abs/0705.4006}{{\tt
  0705.4006}}. [Theor. Math. Phys.155,no.1,567(2008)].

\bibitem{Kazakov:2010iu}
V.~Kazakov, S.~Leurent, and Z.~Tsuboi, ``{Baxter's Q-operators and operatorial
  Backlund flow for quantum (super)-spin chains},'' {\em Commun. Math. Phys.}
  {\bf 311} (2012) 787--814, \href{http://xxx.lanl.gov/abs/1010.4022}{{\tt
  1010.4022}}.

\bibitem{Molev1994}
A.~I. Molev, ``Gelfand-tsetlin basis for representations of yangians,'' {\em
  Letters in Mathematical Physics} {\bf 30} (Jan, 1994) 53--60.

\bibitem{Maillet:2019ayx}
J.~M. Maillet, G.~Niccoli, and L.~Vignoli, ``{Separation of variables bases for
  integrable $gl_{\mathcal{M}|\mathcal{N}}$ and Hubbard models},''
  \href{http://xxx.lanl.gov/abs/1907.08124}{{\tt 1907.08124}}.

\bibitem{2013arXiv1303.1578M}
E.~{Mukhin}, V.~{Tarasov}, and A.~{Varchenko}, ``{Spaces of quasi-exponentials
  and representations of the Yangian Y(gl\_N)},'' {\em ArXiv e-prints} (Mar.,
  2013) \href{http://xxx.lanl.gov/abs/1303.1578}{{\tt 1303.1578}}.

\bibitem{Chernyak:2020lgw}
D.~Chernyak, S.~Leurent, and D.~Volin, ``{Completeness of Wronskian Bethe
  equations for rational gl(m|n) spin chains},''
  \href{http://xxx.lanl.gov/abs/2004.02865}{{\tt 2004.02865}}.

\bibitem{Talalaev:2004qi}
D.~Talalaev, ``{Quantization of the Gaudin system},''
  \href{http://xxx.lanl.gov/abs/hep-th/0404153}{{\tt hep-th/0404153}}.

\bibitem{Zabrodin:1996vm}
A.~Zabrodin, ``{Discrete Hirota's equation in quantum integrable models},''
  {\em Int. J. Mod. Phys.} {\bf B11} (1997) 3125,
  \href{http://xxx.lanl.gov/abs/hep-th/9610039}{{\tt hep-th/9610039}}.

\bibitem{Bazhanov:1989yk}
V.~Bazhanov and N.~Reshetikhin, ``{Restricted Solid on Solid Models Connected
  With Simply Based Algebras and Conformal Field Theory},'' {\em J. Phys.} {\bf
  A23} (1990) 1477.

\bibitem{Cherednik}
I.~Cherednik, ``An analogue of the character formula for hekke algebras,'' {\em
  Funct Anal Its Appl} {\bf 21} (1987) 172–174.

\bibitem{babelon_bernard_talon_2003}
O.~Babelon, D.~Bernard, and M.~Talon, {\em Introduction to Classical Integrable
  Systems}.
\newblock Cambridge Monographs on Mathematical Physics. Cambridge University
  Press, 2003.

\bibitem{Scott:1994dz}
D.~R.~D. Scott, ``{Classical functional Bethe ansatz for SL(N): Separation of
  variables for the magnetic chain},'' {\em J. Math. Phys.} {\bf 35} (1994)
  5831--5843, \href{http://xxx.lanl.gov/abs/hep-th/9403030}{{\tt
  hep-th/9403030}}.

\bibitem{gekhtman1995}
M.~I. Gekhtman, ``Separation of variables in the classical ${\rm sl}(n)$
  magnetic chain,'' {\em Comm. Math. Phys.} {\bf 167} (1995), no.~3 593--605.

\bibitem{Chervov:2007bb}
A.~Chervov and G.~Falqui, ``{Manin matrices and Talalaev's formula},'' {\em J.
  Phys.} {\bf A41} (2008) 194006, \href{http://xxx.lanl.gov/abs/0711.2236}{{\tt
  0711.2236}}.

\bibitem{molev2007yangians}
A.~Molev, {\em Yangians and classical Lie algebras}.
\newblock No.~143. American Mathematical Soc., 2007.

\bibitem{Kuniba:1994na}
A.~Kuniba and J.~Suzuki, ``{Analytic Bethe Ansatz for fundamental
  representations of Yangians},'' {\em Commun. Math. Phys.} {\bf 173} (1995)
  225--264, \href{http://xxx.lanl.gov/abs/hep-th/9406180}{{\tt
  hep-th/9406180}}.

\bibitem{Tsuboi:1997iq}
Z.~Tsuboi, ``{Analytic Bethe ansatz and functional equations for Lie
  superalgebra $sl(r+1|s+1)$},'' {\em J. Phys.} {\bf A30} (1997) 7975--7991,
  \href{http://xxx.lanl.gov/abs/0911.5386}{{\tt 0911.5386}}.

\bibitem{Tsuboi:1998ne}
Z.~Tsuboi, ``{Analytic Bethe Ansatz And Functional Equations Associated With
  Any Simple Root Systems Of The Lie Superalgebra $sl(r+1|s+1)$},'' {\em
  Physica} {\bf A252} (1998) 565--585,
  \href{http://xxx.lanl.gov/abs/0911.5387}{{\tt 0911.5387}}.

\bibitem{Tsuboi:2009ud}
Z.~Tsuboi, ``{Solutions of the T-system and Baxter equations for supersymmetric
  spin chains},'' {\em Nucl. Phys.} {\bf B826} (2010) 399--455,
  \href{http://xxx.lanl.gov/abs/0906.2039}{{\tt 0906.2039}}.

\bibitem{Bazhanov:2010jq}
V.~V. Bazhanov, R.~Frassek, T.~Lukowski, C.~Meneghelli, and M.~Staudacher,
  ``{Baxter Q-Operators and Representations of Yangians},'' {\em Nucl. Phys.}
  {\bf B850} (2011) 148--174, \href{http://xxx.lanl.gov/abs/1010.3699}{{\tt
  1010.3699}}.

\bibitem{Frassek:2011aa}
R.~Frassek, T.~Lukowski, C.~Meneghelli, and M.~Staudacher, ``{Baxter Operators
  and Hamiltonians for 'nearly all' Integrable Closed $\mathfrak{gl}(n)$ Spin
  Chains},'' {\em Nucl. Phys.} {\bf B874} (2013) 620--646,
  \href{http://xxx.lanl.gov/abs/1112.3600}{{\tt 1112.3600}}.

\bibitem{Bazhanov:1996dr}
V.~V. Bazhanov, S.~L. Lukyanov, and A.~B. Zamolodchikov, ``{Integrable
  structure of conformal field theory. 2. Q operator and DDV equation},'' {\em
  Commun. Math. Phys.} {\bf 190} (1997) 247--278,
  \href{http://xxx.lanl.gov/abs/hep-th/9604044}{{\tt hep-th/9604044}}.

\bibitem{Derkachov:2003qb}
S.~E. Derkachov, G.~P. Korchemsky, and A.~N. Manashov, ``{Baxter Q operator and
  separation of variables for the open SL(2,R) spin chain},'' {\em JHEP} {\bf
  10} (2003) 053, \href{http://xxx.lanl.gov/abs/hep-th/0309144}{{\tt
  hep-th/0309144}}.

\bibitem{Niccoli:2010sh}
G.~Niccoli, ``{Reconstruction of Baxter Q-operator from Sklyanin SOV for cyclic
  representations of integrable quantum models},'' {\em Nucl. Phys.} {\bf B835}
  (2010) 263--283, \href{http://xxx.lanl.gov/abs/1001.0035}{{\tt 1001.0035}}.

\bibitem{doi:10.1143/JPSJ.45.321}
R.~Hirota, ``Nonlinear partial difference equations. iv. bäcklund
  transformation for the discrete-time toda equation,'' {\em Journal of the
  Physical Society of Japan} {\bf 45} (1978), no.~1 321--332,
  \href{http://xxx.lanl.gov/abs/https://doi.org/10.1143/JPSJ.45.321}{{\tt
  https://doi.org/10.1143/JPSJ.45.321}}.

\bibitem{Liashyk:2018qfc}
A.~Liashyk and N.~A. Slavnov, ``{On Bethe vectors in
  $\mathfrak{gl}_3$-invariant integrable models},'' {\em JHEP} {\bf 06} (2018)
  018, \href{http://xxx.lanl.gov/abs/1803.07628}{{\tt 1803.07628}}.

\bibitem{Kulish:1983rd}
P.~P. Kulish and N.~{\relax Yu}. Reshetikhin, ``{Diagonalization of GL(N)
  invariant transfer matricies and quantum N wave system (Lee Model)},'' {\em
  J. Phys.} {\bf A16} (1983) L591--L596.

\bibitem{Niccoli:2009jq}
G.~Niccoli and J.~Teschner, ``{The Sine-Gordon model revisited I},'' {\em J.
  Stat. Mech.} {\bf 1009} (2010) P09014,
  \href{http://xxx.lanl.gov/abs/0910.3173}{{\tt 0910.3173}}.

\bibitem{Niccoli:2011nj}
G.~Niccoli, ``{Completeness of Bethe Ansatz by Sklyanin SOV for Cyclic
  Representations of Integrable Quantum Models},'' {\em JHEP} {\bf 03} (2011)
  123, \href{http://xxx.lanl.gov/abs/1102.1694}{{\tt 1102.1694}}.

\bibitem{Marboe:2016yyn}
C.~Marboe and D.~Volin, ``{Fast analytic solver of rational Bethe equations},''
  {\em J. Phys.} {\bf A50} (2017), no.~20 204002,
  \href{http://xxx.lanl.gov/abs/1608.06504}{{\tt 1608.06504}}.

\bibitem{LRV}
S.~Leurent, P.~Ryan, and D.~Volin, ``in preparation,''.

\bibitem{Kazama:2013rya}
Y.~Kazama, S.~Komatsu, and T.~Nishimura, ``{A new integral representation for
  the scalar products of Bethe states for the XXX spin chain},'' {\em JHEP}
  {\bf 09} (2013) 013, \href{http://xxx.lanl.gov/abs/1304.5011}{{\tt
  1304.5011}}.

\bibitem{Kitanine:2015jna}
N.~Kitanine, J.~M. Maillet, G.~Niccoli, and V.~Terras, ``{On determinant
  representations of scalar products and form factors in the SoV approach: the
  XXX case},'' {\em J. Phys.} {\bf A49} (2016), no.~10 104002,
  \href{http://xxx.lanl.gov/abs/1506.02630}{{\tt 1506.02630}}.

\bibitem{Gromov:2019wmz}
N.~Gromov, F.~Levkovich-Maslyuk, P.~Ryan, and D.~Volin, ``{Dual Separated
  Variables and Scalar Products},''
  \href{http://xxx.lanl.gov/abs/1910.13442}{{\tt 1910.13442}}.

\bibitem{Cavaglia:2019pow}
A.~Cavaglià, N.~Gromov, and F.~Levkovich-Maslyuk, ``{Separation of variables
  and scalar products at any rank},''
  \href{http://xxx.lanl.gov/abs/1907.03788}{{\tt 1907.03788}}.

\bibitem{Gromov:2020fwh}
N.~Gromov, F.~Levkovich-Maslyuk, and P.~Ryan, ``{Determinant Form of
  Correlators in High Rank Integrable Spin Chains via Separation of
  Variables},'' \href{http://xxx.lanl.gov/abs/2011.08229}{{\tt 2011.08229}}.

\bibitem{Gromov:2018cvh}
N.~Gromov and F.~Levkovich-Maslyuk, ``{New Compact Construction of Eigenstates
  for Supersymmetric Spin Chains},'' {\em JHEP} {\bf 09} (2018) 085,
  \href{http://xxx.lanl.gov/abs/1805.03927}{{\tt 1805.03927}}.

\bibitem{Derkachov:2018ewi}
S.~E. Derkachov and P.~A. Valinevich, ``{Separation of variables for the
  quantum $SL(3,\mathbb C)$ spin magnet: eigenfunctions of Sklyanin
  $B$-operator},'' \href{http://xxx.lanl.gov/abs/1807.00302}{{\tt 1807.00302}}.

\bibitem{Valinevich:2016cwq}
P.~A. Valinevich, S.~Derkachov, P.~P. Kulish, and E.~M. Uvarov, ``{Construction
  of eigenfunctions for a system of quantum minors of the monodromy matrix for
  an $SL(n,\mathbb C)$-invariant spin chain},'' {\em Theor. Math. Phys.} {\bf
  189} (2016), no.~2 1529--1553. [Teor. Mat. Fiz.189,no.2,149(2016)].

\bibitem{ValinevichGT}
P.~A. Valinevich, ``{Construction of the Gelfand–Tsetlin Basis for Unitary
  Principal Series Representations of the Algebra $sl(n,\mathbb{C})$.},'' {\em
  Theor. Math. Phys.} {\bf 198} (2019) 145--155.

\bibitem{Gurdogan:2015csr}
O.~Gurdogan and V.~Kazakov, ``{New Integrable 4D Quantum Field Theories from
  Strongly Deformed Planar $\mathcal N = $ 4 Supersymmetric Yang-Mills
  Theory},'' {\em Phys. Rev. Lett.} {\bf 117} (2016), no.~20 201602,
  \href{http://xxx.lanl.gov/abs/1512.06704}{{\tt 1512.06704}}. [Addendum: Phys.
  Rev. Lett.117,no.25,259903(2016)].

\bibitem{Kazakov:2018qez}
V.~Kazakov and E.~Olivucci, ``{Biscalar Integrable Conformal Field Theories in
  Any Dimension},'' {\em Phys. Rev. Lett.} {\bf 121} (2018), no.~13 131601,
  \href{http://xxx.lanl.gov/abs/1801.09844}{{\tt 1801.09844}}.

\bibitem{Maillet:2018rto}
J.~M. Maillet and G.~Niccoli, ``{Complete spectrum of quantum integrable
  lattice models associated to $\mathcal{U}_{q} (\widehat{gl_{n}})$ by
  separation of variables},'' {\em J. Phys.} {\bf A52} (2019), no.~31 315203,
  \href{http://xxx.lanl.gov/abs/1811.08405}{{\tt 1811.08405}}.

\end{thebibliography}\endgroup
\end{document}